\documentclass{article}

\usepackage{microtype}
\usepackage{graphicx}
\usepackage{subfigure}
\usepackage{booktabs} %

\usepackage{quoting}
\quotingsetup{leftmargin=0.2cm,rightmargin=0.2cm,vskip=5pt}

\usepackage{hyperref}

\usepackage[accepted]{icml2025}
\usepackage{natbib}
\usepackage{enumitem}
\usepackage[labelfont=bf]{caption}
\usepackage{makecell}
\usepackage{listings}
\usepackage{hyperref}

\usepackage{amsmath}
\usepackage{amssymb}
\usepackage{mathtools}
\usepackage{amsthm}

\usepackage[capitalize,noabbrev]{cleveref}

\theoremstyle{plain}

\theoremstyle{definition}

\theoremstyle{remark}

\usepackage[textsize=tiny]{todonotes}
\usepackage{array}
\usepackage{rotating} %
\usepackage{longtable} %
\newcommand{\centeredrowheader}[1]{\parbox[c][0.4cm][c]{2.3cm}{#1}}

\newcommand{\fancymark}{\raisebox{\dimexpr-0.5\height+0.5\depth\relax}{\ensuremath{\checkmark}}}

\newcommand{\squarecell}[1]{\makebox[0.4cm]{\rule[-0.2cm]{0pt}{0.4cm}#1}}

\icmltitlerunning{Infected Smallville: How Disease Threat Shapes Sociality in LLM Agents}

\begin{document}

\twocolumn[
\icmltitle{Infected Smallville: How Disease Threat Shapes Sociality in LLM Agents}

\icmlsetsymbol{equal}{*}

\begin{icmlauthorlist}
\icmlauthor{Soyeon Choi}{yyy}
\icmlauthor{Kangwook Lee}{yyyy}
\icmlauthor{Oliver Sng}{zzz}
\icmlauthor{Joshua M. Ackerman}{xxx}
\end{icmlauthorlist}

\icmlaffiliation{yyy}{Department of Psychology, University of Wisconsin-Madison}
\icmlaffiliation{yyyy}{Department of Electrical and Computer Engineering, University of Wisconsin-Madison}
\icmlaffiliation{zzz}{Department of Psychological Science, University of California, Irvine}
\icmlaffiliation{xxx}{Department of Psychology, University of Michigan, Ann Arbor}

\icmlcorrespondingauthor{Soyeon Choi}{soyeon.choi@wisc.edu}

\icmlkeywords{LLM agents, psychology of LLM agents, AI for social psychology, infectious disease, sociality, behavioral immune system, ICML}

\vskip 0.3in
]

\printAffiliationsAndNotice{}

\begin{abstract}
How does the threat of infectious disease influence sociality among generative agents? We used generative agent-based modeling (GABM), powered by large language models, to experimentally test hypotheses about the behavioral immune system. Across three simulation runs, generative agents who read news about an infectious disease outbreak showed significantly reduced social engagement compared to agents who received no such news, including lower attendance at a social gathering, fewer visits to third places (e.g., cafe, store, park), and fewer conversations throughout the town. In interview responses, agents explicitly attributed their behavioral changes to disease-avoidance motivations. A validity check further indicated that they could distinguish between infectious and noninfectious diseases, selectively reducing social engagement only when there was a risk of infection. Our findings highlight the potential of GABM as an experimental tool for exploring complex human social dynamics at scale.
\end{abstract}

\vspace{-0.2in}
\section{Introduction}
\subsection{Generative Agents for Social Psychology}
Recent developments have sparked interest in integrating generative agents, which leverage large language models (LLMs), into psychological research. Although these generative agents are not equivalent to human participants, their reliance on linguistic processes and social interactions analogous to human behaviors allows them to replicate various aspects of social cognition and behavior \cite{anthis2025llm,argyle2023out,demszky2023using,mei2024turing,vezhnevets2023generative}. For example, generative agents created from qualitative interview data replicated responses from the General Social Survey with approximately 85\% accuracy, comparable to the consistency of participants' responses over time, while also reducing demographic biases \cite{park2024generative}. Beyond survey replication, generative agents have demonstrated the capacity for emergent social behaviors at the group level, including conformity \cite{zhang2023exploring}, opinion dynamics \cite{chuang2023simulating}, cooperation \cite{park2023generative,kaiya2023lyfe}, negotiation \cite{zhou2023sotopia}, and collaborative task performance \cite{qian-etal-2024-chatdev}. As such, generative agent-based modeling (GABM) offers a novel tool for testing classic and contemporary theories in social psychology.

Topics within social psychology that might particularly benefit from implementing GABM include those examining how ecological dimensions (macroenvironmental factors such as population density, genetic relatedness, sex ratio, resources, mortality likelihood, and pathogen threat) influence social cognition and behavior \cite{gelfand2011differences,neuberg2011human,oishi2010social,sng2018behavioral}. Given the practical and ethical challenges of manipulating ecological conditions or perceptions in real-world settings over extended periods, GABM provides a viable methodological alternative for systematically testing theoretical predictions. A similar recommendation, to simulate ecological conditions and animal interactions, has been proposed in behavioral ecology, a discipline that shares many topical overlaps \cite{rafiq2025generative}. In the current paper, we specifically examine the dimension of pathogen threat, focusing on how manipulating perceptions of infectious disease risk influences sociality among generative agents over a two-day period.

\begin{figure*}[ht!]
\vspace{-0.05in}
    \centering
\includegraphics[width=0.75\textwidth]{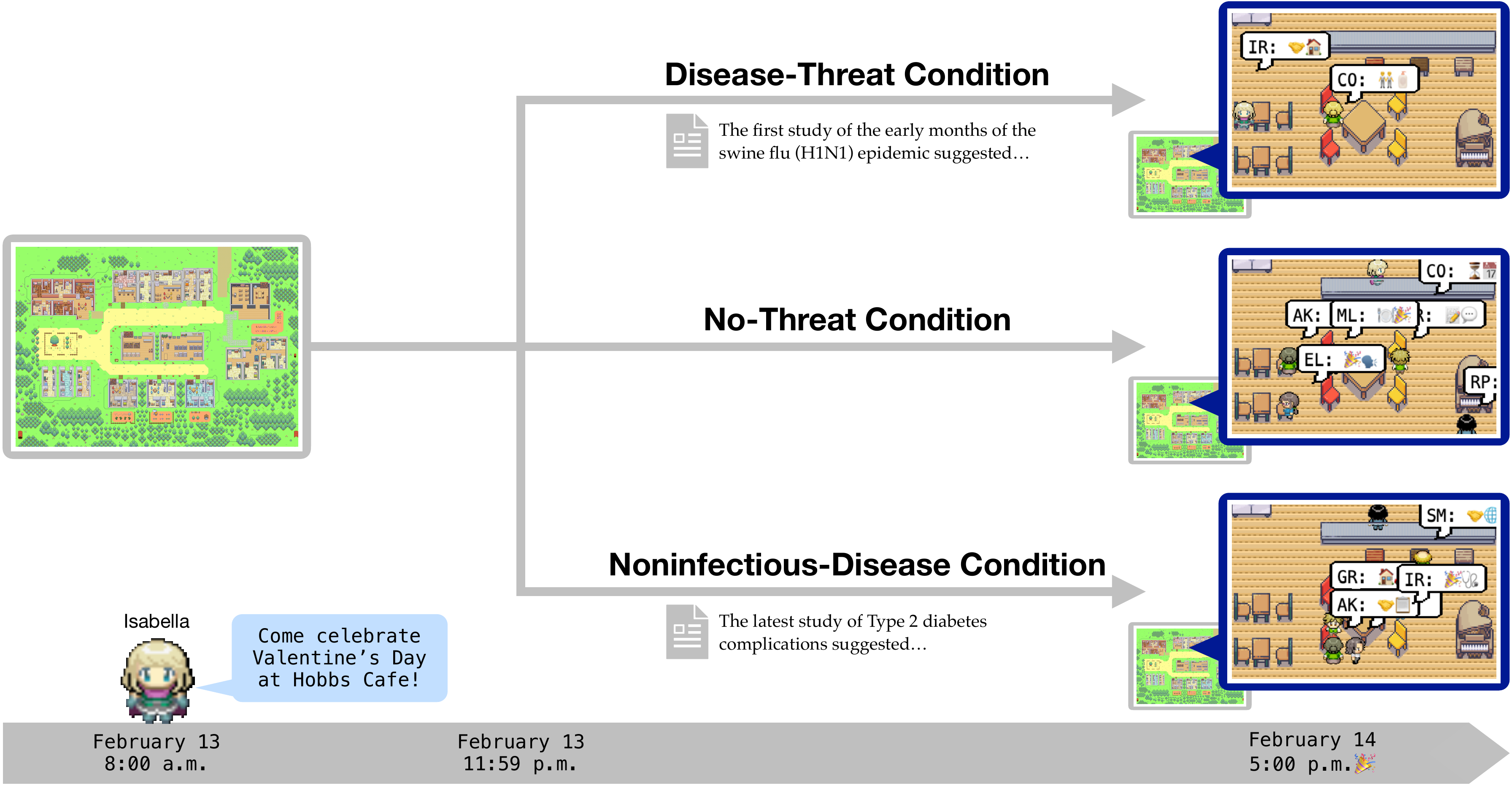}
    \caption{\textit{Overview of Experimental Design.} Experimental conditions began at 11:59 p.m. on February 13, creating trajectories analogous to a multiverse structure. The noninfectious-disease condition was included only in the first run.}
    \label{fig:figure1}
    \vspace{-0.15in} %
\end{figure*}

\subsection{Simulating the Behavioral Immune System}
Avoiding infectious diseases has been a fundamental and recurrent adaptive challenge throughout human history \cite{wolfe2007origins}, and will continue to be, as illustrated by recent disease outbreaks such as COVID-19 and Mpox. Psychologists have proposed that in addition to the biological immune system, which functions to combat pathogens already within the body, humans possess a \textit{behavioral immune system}, which functions to preemptively prevent pathogen exposure by detecting and avoiding people or situations that carry a risk of infection \cite{ackerman2018behavioral, schaller2007behavioral}. 
When individuals encounter situational cues indicating pathogen threat, such as reading news about disease outbreaks, being exposed to disgusting images or unpleasant odors, or detecting signs of sickness in others (e.g., coughing, pale appearance), they exhibit avoidant behaviors to minimize the risk of pathogen transmission, particularly in contexts involving social interaction. These behaviors may include spending more time at home, making fewer visits to nonessential locations (e.g., restaurants, department stores), and being more reluctant to interact with unfamiliar individuals \cite{choi2022keep, gao2020mapping, gollwitzer2022linking, mortensen2010infection, sacco2014balancing}.

In testing hypotheses about the behavioral immune system, GABM could serve as a novel tool that complements many existing methodologies and provides unique insights into human social dynamics. Common methods for studying human behavior, including surveys and lab experiments (e.g., using vignettes or images of disease-related stimuli), provide controlled insights but often lack ecological validity as they do not fully capture social cognition and behavior as they naturally occur in everyday life. More naturalistic methods, including daily diary and experience sampling studies, enhance ecological validity but face significant challenges in clearly establishing causation (such as pinpointing epidemic onset) and managing numerous confounding factors inherent to real-world settings. Traditional agent-based modeling partially addresses these challenges but is fundamentally rule-based, and agent interactions involve simple, abstract communication mechanisms \cite{smith2007agent}.

GABM complements these methods by allowing agents to create short- and long-term plans (e.g., daily schedules, event planning), interact with the simulated environment (e.g., entering and leaving buildings, using objects), communicate with other agents (e.g., initiating conversations), and form relationships (e.g., friendships, romantic relationships), all using natural language. These capabilities closely resemble real-world behaviors, enabling complex, nonlinear interactions in response to ecological changes. Recent work by \citet{williams2023epidemic} demonstrated that combining conventional agent-based epidemic modeling with decision-making guided by large language models can better represent realistic human decision-making processes. In their study, agents received natural-language prompts about personal symptoms and daily infection rates, and independently decided whether to quarantine when feeling sick or self-isolate as cases increased. Incorporating these LLM-guided decisions into the conventional model resulted in fewer contacts and a flatter epidemic curve. However, to our knowledge, no research has yet explored how fully leveraging LLMs to simultaneously simulate disease threat perception, individual decision-making, and realistic social interactions influences broader patterns of sociality among agents.

\subsection{Our Contributions}
We integrated empirical tests of behavioral immune system hypotheses with a GABM framework designed to simulate human behavior across varied scenarios \cite{park2023generative}. Following the experimental paradigm from \citet{huang2011immunizing}, we primarily compared two simulated conditions: a disease-threat condition in which agents received information about an ongoing swine flu outbreak via a local newspaper article, and a control condition without such information. This approach allowed us to observe naturally occurring differences in social cognition and behaviors between and within these simulations over time. We hypothesized that agents in the disease-threat condition would show more reserved social behaviors, including reduced physical movement within their environment and fewer interpersonal interactions, reflecting heightened concern about infection risks. We also conducted qualitative interviews with agents to explore their reasoning behind behavioral changes and administered standardized psychological scales, consistent with established psychological research methods. To assess the validity of our findings, we included an additional condition involving a noninfectious disease to examine whether agents selectively modify their social behaviors in response to infectious disease threats (see Figure \ref{fig:figure1}). Across three independent simulation runs, we found that the perceived threat from swine flu significantly reduced agents’ spatial movement and social interactions, as they prioritized disease avoidance over social affiliation.

\section{Method}
\subsection{Sandbox Environment}
We adapted the interactive sandbox environment developed by \citet{park2023generative}, which simulates a small town called Smallville, inhabited by 25 natural language-driven agents. We utilized GPT-4.1-mini for our implementation and GPT-4.1 for interviews \cite{openai2025chatgpt}. The environment closely resembles life simulation games such as The Sims or inZOI \cite{maxis2000sims, krafton2024inzoi}, featuring locations including cafes, bars, parks, schools, dormitories, houses, and stores, all equipped with functional areas and objects. Each agent in Smallville is uniquely identified by a name and an avatar, along with their identity, occupation, lifestyle, personality traits, and relationship histories. For example, Latoya Williams, a 25-year-old digital photographer, is described as organized, logical, and attentive. She lives in an artists' co-living space with four housemates and is currently working on photography inspired by her travels.

The agents operate through a cycle: they (1) store experiences in long-term memory, such as their own actions, observations of other agents or objects, and social interactions, (2) reflect on these memories to form higher-level insights about themselves and their environment, (3) plan daily activities based on these reflections, and (4) take actions expressed through natural language statements, emojis describing their current activities, and movements across the map. This cycle repeats continuously, as each action feeds back into the agents’ memory streams and influences their subsequent behaviors. We refer readers to the original study \cite{park2023generative} for a detailed description of the agent architecture.

While we followed most of the system design, we made a few changes that enhance the robustness of the system. First, we used structured output for every LLM inference, significantly reducing the likelihood of invalid inference outputs. Without structured output, we observed that LLM inferences frequently made mistakes, for instance defaulting to hardcoded failsafe options such as visiting the park. Structured output allowed us to enforce model constraints effectively. Second, we incorporated chain-of-thought prompting into almost every LLM inference within the agent system. This addition notably increased inference latency and cost, but it also substantially improved the fidelity of individual inferences. Third, we modified the approach agents use to decide whether to initiate conversations. The original approach, which relied on a direct binary decision (true or false) from the LLM, was overly sensitive to minor variations in prompts. To address this, we introduced a scoring mechanism in which the LLM assigns a likelihood score for initiating conversations. This score was then converted into a probability and used to probabilistically determine whether a conversation is initiated (see Sec. 2.3.2 for details).

We also modified the original interview implementation, which previously retrieved a limited set of relevant memory entries using sentence embeddings, summarized them, and combined the summary with interview questions for the LLM. This earlier approach often omitted critical information and produced inaccuracies. Given recent advancements in LLM context length and long-context processing, we eliminated the summarization step and instead directly concatenated up to 1000 relevant raw memory entries with interview questions. This change substantially improved the accuracy of the generated responses.

In \citet{park2023generative}, a notable event began with Isabella Rodriguez, owner of Hobbs Cafe who decided to host a Valentine’s Day party. Information about the party soon spread throughout the town as agents autonomously invited others, formed new relationships, and coordinated their attendance. This emergent coordination illustrated the agents’ ability to simulate realistic and complex social behaviors. The current research specifically examines the social dynamics surrounding this key event in Smallville, particularly focusing on spatial behaviors and social interactions.

\subsection{Experimental Design}
We conducted simulations using a 2 (condition: disease-threat vs. no-threat) $\times$ 2 (day: February 13 vs. February 14) design. Condition served as a between-simulations factor, whereas day was a repeated-measures factor within each simulation. Analogous to a multiverse structure, both conditions began identically on February 13 and diverged into either the disease-threat or no-threat condition on February 14. For robustness, we ran three independent pairs of simulations, each consisting of one disease-threat and one no-threat condition, resulting in a total of six simulations (see Figure \ref{fig:figure1}).

In the disease-threat condition, agents read a news article about an ongoing swine flu (H1N1) epidemic (adapted from \citealp{huang2011immunizing}) before going to bed on February 13. They were asked to reflect on this information and then plan activities for the next day. Agents in the no-threat condition did not receive any news article and proceeded directly to planning activities for the following day (see Sec.~\ref{sec:app_a} for the full prompt).

\vspace{-0.1in}
\begin{quoting}
\small
\emph{The first study of the early months of the swine flu (H1N1) epidemic suggested that one quarter of people in our country sick enough to be hospitalized with swine flu last spring also wound up needing intensive care; of these, 7 percent of them died. These rates are higher than with ordinary seasonal flu, and experts note that what is striking and unusual is that healthy people accounted for nearly half of the hospitalized cases. In total, health authorities claim that over 2 million people across the country could be hospitalized because of swine flu. They also noted that individuals might be infectious even while asymptomatic, increasing the risk of unnoticed transmission in our community.}
\end{quoting}

\subsection{Measures}
Sociality was assessed through two quantitative aspects (agents’ spatial behavior and social interaction) and two qualitative methods (interview and case study).

\subsubsection{Spatial Behavior}
\textit{Attendance at Valentine’s Day Party.} We measured how many agents attended the party at Hobbs cafe (excluding the cafe owner, Isabella Rodriguez) and how long each agent stayed between 5 p.m. and 7 p.m. on February 14. Higher attendance and longer durations indicate greater social engagement among agents.

\textit{Time Spent in Third Places.} We recorded the amount of time agents spent in third places in Smallville, including a cafe, bar, park, supply store, and grocery and pharmacy (see Sec.~\ref{sec:app_a}). Third places are informal public spaces outside of home (first place) and work (second place), where people casually come together, interact, and foster a sense of community \cite{oldenburg1989great}. More time spent in third places reflects higher levels of interest in voluntary social participation and community engagement; we classified these locations as third places based on their social function, regardless of whether other agents were present at any specific moment. We also measured the number of steps taken by agents as a supplementary indicator of overall mobility.

\subsubsection{Social Interaction}
\textit{Conversational Patterns: Initiation, Frequency, and Length.}
Whenever an agent encountered another agent and this encounter satisfied certain conditions (e.g., proximity, no recent interaction), the agent made an LLM inference—reflecting their status, memory, and prior interaction history with the other agent—to assign a conversation initiation score, where 10 indicates it is highly likely and natural to initiate a conversation, 0 indicates absolute unlikelihood, and 5 represents neutrality. We then converted this score into a probability, denoted by $p = 0.1 \times \text{score}$. For the $i$-th encounter, let $p_i$ be the estimated probability from the LLM inference, and let $X_i \sim \text{Bernoulli}(p_i)$ represent the actual binary outcome of whether a conversation is initiated. Suppose there are $n$ encounters. Then,
\begin{align*}
\text{Average conversation initiation probability} &= \textstyle\frac{1}{n}\textstyle\sum_{i=1}^{n} p_i \\[5pt]
\text{Encounter-to-conversation frequency} &= \textstyle\frac{1}{n}\textstyle\sum_{i=1}^{n} X_i
\end{align*}
Since \(E[X_i] = p_i\), it follows that the expected encounter-to-conversation frequency equals the average conversation initiation probability. We report the encounter-to-conversation frequency in the main body and defer the average conversation initiation probability to Sec.~\ref{sec:app_b}.

For conversations that were initiated, we recorded the total number of conversations across the town and the number of turns within each conversation (up to a maximum of eight turns per conversation). Higher initiation probabilities, conversation frequencies, and number of turns per interaction all indicate greater social engagement. We also report these metrics separately for interactions with familiar versus unfamiliar agents, based on the default relationship history established at the beginning of February 13 (see Sec.~\ref{sec:app_a} for the relationship mapping).

\textit{Information Diffusion.} We counted how many agents were aware of the Valentine’s Day party five minutes prior to its start (at 4:55 p.m. on February 14) by manually inspecting their memory. Higher numbers of informed agents suggest wider information dissemination and stronger social connectivity among agents.

\subsubsection{Interview}
Agents in both conditions participated in structured interviews at the end of the day on February 14, reflecting on their social activities and interactions.

\textit{Responses to Psychological Scale.} Agents completed a standardized scale, as would human participants in social psychology studies. They responded to selected items from the Fundamental Social Motives Inventory \cite{neel2016individual}, focusing on two motives: Disease Avoidance and Social Affiliation. The Disease Avoidance scale includes items such as “I avoid places and people that might carry diseases.” The Social Affiliation scale consists of three distinct submotives: Exclusion Concern (e.g., “I would be extremely hurt if a friend excluded me”), Group (e.g., “Being part of a group is important to me”), and Independence (e.g., “I prefer to spend time alone than to be surrounded by other people”). All items were rated on a 7-point scale (1 = \textit{strongly disagree}, 7 = \textit{strongly agree}).

\textit{Responses to Interview Questions.}  Agents also answered interview questions in an open-ended format. These interviews allowed us to examine the underlying motivations behind their social behaviors. The questions were as follows:

\begin{itemize}[topsep=0.15em,itemsep=0.15em,parsep=0pt,partopsep=0pt]
    \item Did you attend the Valentine’s Day party at Hobb’s Cafe today? If yes, please explain why you attended. If no, please explain why you did not attend.
    \item \textit{[Disease-threat condition]} How do you feel about the current swine flu (H1N1) situation? Is this a topic of concern for you?
    \item \textit{[Disease-threat condition]} Have you taken or planned any specific actions to protect yourself or others from swine flu (H1N1)?
\end{itemize}

\subsubsection{Case Study}
As a qualitative illustration of how swine flu news influenced agents in the disease-threat condition compared to those in the no-threat condition, we conducted a case study of Isabella Rodriguez, owner of Hobbs Cafe. We analyzed logs documenting Isabella’s thought processes and activities during her preparations for the Valentine's Day party.

\subsection{Validity Check: H1N1 Flu vs. Type 2 Diabetes}
For the simulation to be valid, it is important that the threat of infectious diseases exerts a unique effect on predicting sociality. Similar methodological approaches have been adopted in the behavioral immune system literature, where researchers sometimes use a comparison condition involving threats unrelated to infectious diseases, such as car accidents, tornadoes, or active shooters, rather than simply a no-threat control (e.g., \citealp{faulkner2004evolved, wang2019infectiousness}). The idea is that if the predicted behavioral responses emerge exclusively under the infectious disease threat condition and not under conditions involving other types of threats, it would indicate that these responses are uniquely driven by concerns about infectious disease rather than by a general threat perception.

Our simulation extends this approach by incorporating a closely related yet distinct condition: a noninfectious disease. Specifically, in the first run of our simulation, we introduced an additional condition in which agents read an article about increased hospitalizations due to complications from type 2 diabetes, instead of one about swine flu (see Sec.~\ref{sec:app_a}).

If the outcomes in the noninfectious-disease condition parallel those observed in the no-threat condition more closely than the infectious disease-threat condition (e.g., a successful Valentine’s Day party), this would indicate that agents effectively distinguish between different types of threats, adjusting their social behavior specifically in response to infectious (but not noninfectious) disease risks.

\section{Results}
Outcomes reported are presented either as town-level totals or as agent-level means, depending on what is most meaningful for each measure. Town-level totals were computed by summing the values across all 25 agents within each run, and then averaging these totals across the three runs within each condition. Agent-level means were computed by first calculating the mean outcome across the 25 agents within each run, and then (macro-)averaging these means across the three runs within each condition. We focused on comparing outcomes between the disease-threat and no-threat conditions on February 14. Where relevant, we also report changes from February 13 to 14 and how these changes differed across conditions. Detailed results for each run on February 13 and 14 are provided in Sec.~\ref{sec:app_b}.

\subsection{Spatial Behavior}
\subsubsection{Attendance at Valentine’s Day Party}
\begin{figure}[H]
  \vspace{-0.05in}
  \centering
  \includegraphics[width=\columnwidth]{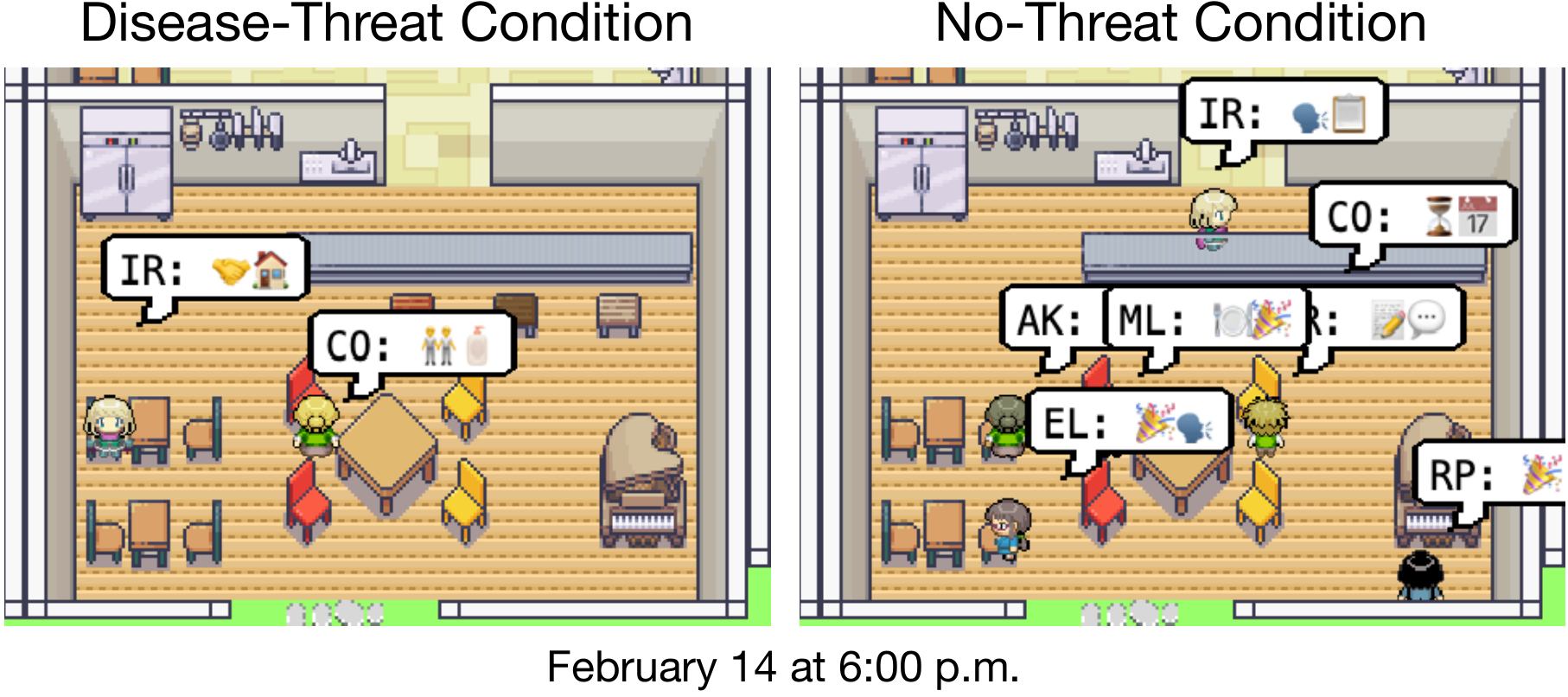}
  \caption{\textit{Valentine's Day party at Hobbs Cafe.} In the first run, only one agent attended the party in the disease-threat condition, whereas eight agents attended in the no-threat condition. See Sec.~\ref{sec:app_b} for screenshots from other runs.}
  \label{fig:figure2}
  \vspace{-0.225in}
\end{figure}

We observed the number of agents, out of a total of 25 from Smallville, who visited Hobbs Cafe between 5 p.m. and 7 p.m. on February 14. In the disease-threat condition, an average of 1.33 agents were present at the cafe across the three runs. However, closer examination revealed substantial variation across these runs. In the first run, the party proceeded as planned, but only one agent attended (see Figure \ref{fig:figure2}).
During the second and third runs, cafe owner Isabella Rodriguez postponed the Valentine’s Day party. Despite the postponement, two agents visited independently in the second run, while in the third run, one agent arrived unaware of the postponement and was subsequently informed by Isabella.

In the no-threat condition, the party was held as scheduled, with attendance averaging 7.33 agents per run (see Sec.~\ref{sec:app_b} for the attendee list). The number of attendees is similar to the five agents reported by \citet{park2023generative}. Those who attended remained at the cafe for an average duration of 80.09 minutes.

Thus, the news about swine flu clearly reduced social engagement. For example, Maria Lopez, a 20-year-old student at Oak Hill College, initially planned to attend the party during the first run of the disease-threat condition. However, after reading the news, she decided to stay home to study physics. In contrast, in the no-threat condition, Maria attended as originally planned.

\subsubsection{Time Spent in Third Places}
\begin{figure}[h]
  \vspace{0.1in}
  \centering
  \includegraphics[width=\columnwidth]{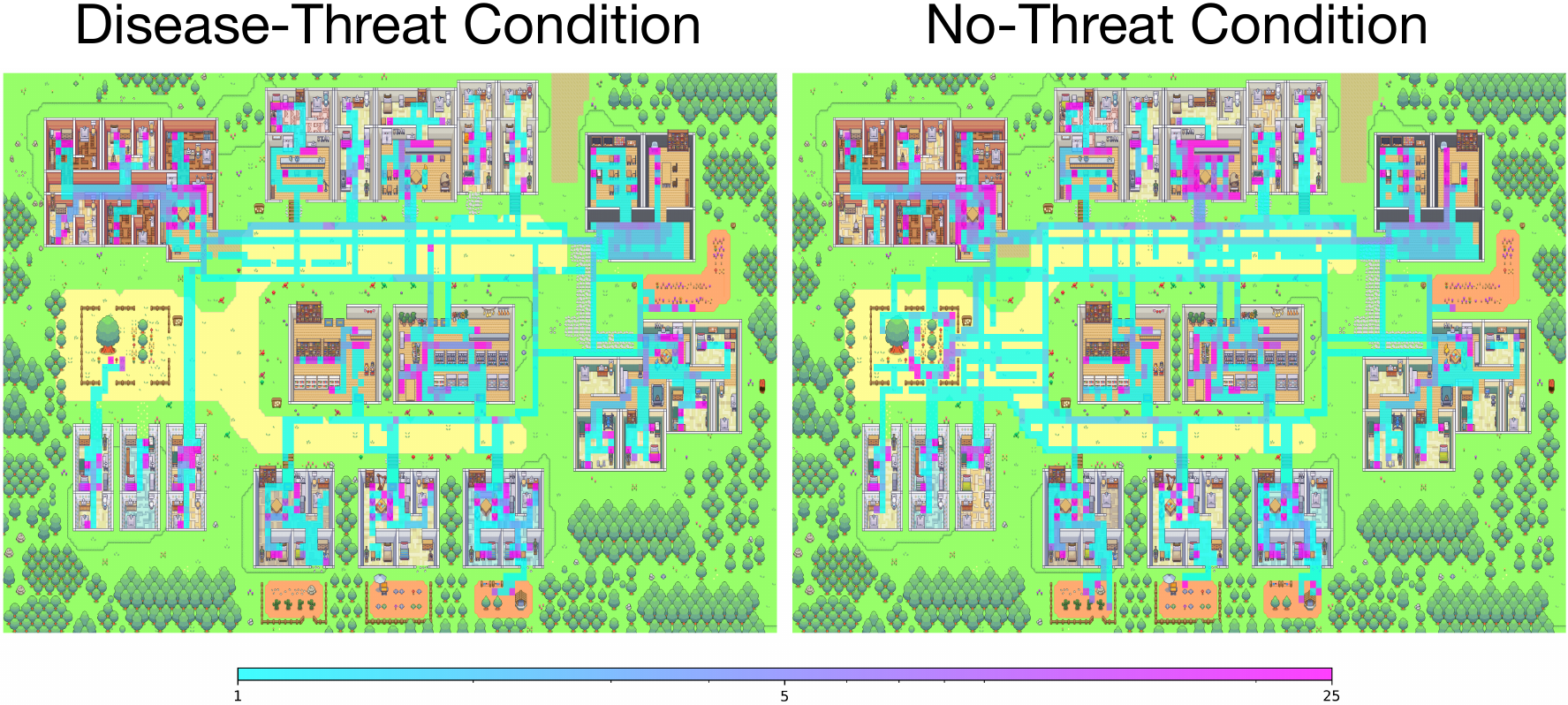}
  \caption{\textit{Heatmap of Smallville.} The heatmap displays the total time (min) summed across all agents in the third run, with pink shades indicating more time spent and blue shades indicating less. See Sec.~\ref{sec:app_b} for heatmaps of other runs.}
  \label{fig:figure3}
  \vspace{-0.1in}
\end{figure}

In the disease-threat condition, agents spent 73.3\% less time in third places (locations other than home or work) on February 14 (30.11 minutes) compared to the no-threat condition (112.65 minutes). They also exhibited lower mobility, taking on average 22.3\% fewer steps (437.25 steps) compared to agents in the no-threat condition (562.69 steps). See Figure \ref{fig:figure3} for a heatmap visualization of these differences.

\vspace{0.1in}

When examining changes from February 13 to February 14, agents in the disease-threat condition showed a 64.2\% decrease in time spent in third places (from 84.20 to 30.11 minutes) and 14.9\% fewer daily steps (from 513.93 to 437.25 steps). In contrast, agents in the no-threat condition showed a 33.8\% increase in time spent in third places (from 84.20 to 112.65 minutes) and 9.5\% more daily steps (from 513.9 to 562.69 steps) over the same period. These findings suggest that the threat of infectious disease not only reduced interest in social engagement but also corresponded with decreased physical movement within the community.

\subsection{Social Interaction}
\subsubsection{Conversational Patterns}
In the disease-threat condition, agents were 12.3 percentage points less likely to initiate conversations (49.9\%) compared to the no-threat condition (62.2\%). Similarly, they engaged in 48.4\% fewer conversations overall in the disease-threat condition (73.67 total conversations) relative to the no-threat condition (142.67 total conversations). There was little difference in conversation length between the disease-threat (5.51 turns) and no-threat (5.68 turns) conditions.

From February 13 to February 14, agents in the disease-threat condition showed notable decreases in their probability of initiating conversations (12.3 percentage points, from 62.20\% to 49.9\%) and total conversations (33.2\% decrease, from 110.33 to 73.67 total conversations). Conversation length decreased slightly (from 5.70 to 5.51 turns). In contrast, agents in the no-threat condition maintained essentially the same probability of initiating conversations (0.03 percentage point increase, from 62.20\% to 62.23\%), engaged in more conversations (29.3\% increase, from 110.33 to 142.67 total conversations; likely due to the party), and had minimal change in conversation length (from 5.70 to 5.68 turns).

\begin{figure}[h]
  \vspace{0.05in}
  \centering
  \includegraphics[width=\columnwidth]{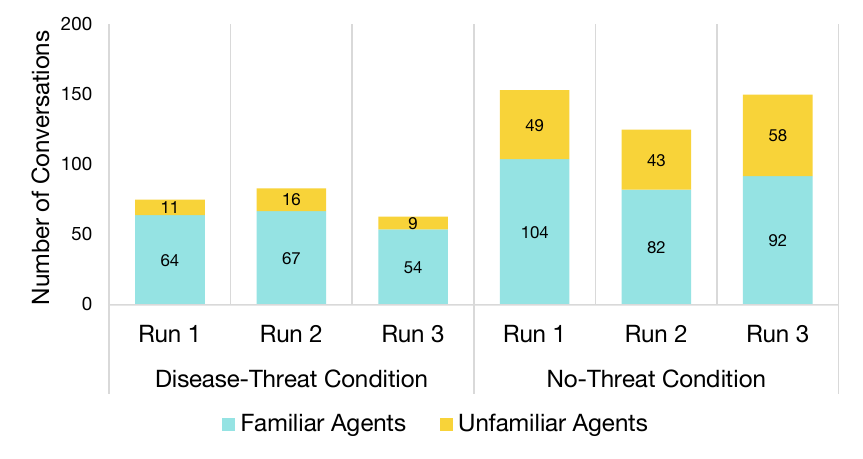}
  \caption{\textit{Number of Conversations Across Smallville.}}
  \label{fig:figure4}
  \vspace{-0.1in}
\end{figure}

Interestingly, the differences between conditions were particularly pronounced in interactions with unfamiliar agents (i.e., those without prior interaction history at the beginning of February 13) compared to familiar agents. Agents in the disease-threat condition were 11.5 percentage points less likely to initiate conversations with familiar agents than those in the no-threat condition, but 23.2 percentage points less likely to do so with unfamiliar agents. Together with lower mobility, this resulted in a significant overall reduction in conversations throughout the town (see Figure \ref{fig:figure4}).

\subsubsection{Information Diffusion}
There was a slight difference in the number of agents aware of the Valentine’s Day party at 4:55 p.m. on February 14, with fewer agents aware in the disease-threat condition (10.67 agents on average) compared to the no-threat condition (12.67 agents on average). Although information dissemination declined slightly, likely resulting from reduced physical movement and fewer interactions, this finding suggests that limited awareness was not the critical factor influencing attendance. Rather, agents in the disease-threat condition who were aware of the party explicitly chose not to attend, suggesting that avoidance of social contact primarily contributed to the lower attendance.

\subsection{Interview}
\subsubsection{Responses to Psychological Scale}
On a 7-point scale, agents in the disease-threat condition reported on average a 2.78-point higher motivation for disease avoidance (6.26) compared to the no-threat condition (3.48), reflecting heightened concerns about swine flu. On the other hand, differences in social affiliation motives were minimal, with agents in the disease-threat condition reporting slightly higher exclusion concern (0.02 points), lower desire for group affiliation (0.19 points), and higher motive for independence from others (0.28 points) compared to the no-threat condition (see Table \ref{tab:fsm_inventory}).

\vspace{0.1in}
\begin{table}[h]
  \vspace*{-0.5\baselineskip}             %
  \centering
  \small                              %
  \setlength{\tabcolsep}{1.75pt}           %
  \renewcommand{\arraystretch}{1.5}       %
  \caption{\textit{Responses to Fundamental Social Motives Scale}}
  \label{tab:fsm_inventory}
  \resizebox{\columnwidth}{!}{%
    \begin{tabular}{lcccccccc}
      \toprule
       & \multicolumn{4}{c}{\textbf{Disease-Threat Condition}}
       & \multicolumn{4}{c}{\textbf{No-Threat Condition}} \\
      \cmidrule(lr){2-5}\cmidrule(lr){6-9}
      Measures 
        & Run 1 & Run 2 & Run 3 & \textit{M} 
        & Run 1 & Run 2 & Run 3 & \textit{M} \\
      \midrule
      Disease Avoidance
        & 6.13 & 6.17 & 6.35 & 6.26
        & 3.43 & 3.52 & 3.49 & 3.48 \\
      Affiliation (Exclusion Concern)
        & 4.07 & 4.03 & 4.08 & 4.06
        & 4.03 & 3.97 & 4.11 & 4.04 \\
      Affiliation (Group)
        & 6.10 & 6.15 & 6.05 & 6.10
        & 6.29 & 6.35 & 6.23 & 6.29 \\
      Affiliation (Independence)
        & 3.41 & 3.27 & 3.49 & 3.39
        & 3.12 & 3.09 & 3.11 & 3.11 \\
      \bottomrule
    \end{tabular}%
  }
  \vspace*{-0.6\baselineskip}               %
\end{table}
\textit{Note}. The items were rated by agents using a 7-point scale (1 = \textit{strongly disagree}; 7 = \textit{strongly agree}).
\vspace{0.05in}

\subsubsection{Responses to Interview Questions}
Following the scale responses, we conducted interviews with the agents. Due to space constraints, the interview responses are presented in Sec.~\ref{sec:app_b}, focusing on the agents who attended the party in the no-threat condition but did not attend in the disease-threat condition.

In the disease-threat condition, agents consistently mentioned concerns about swine flu as their primary reason for not attending the party, and these responses matched their actual behaviors. They emphasized prioritizing health and safety, as well as other responsibilities such as music projects or app development, over social gatherings. To manage their risk, agents took practical steps, including limiting social interactions and maintaining good hygiene. For example, in the first run, several agents decided to help with party preparations but did not attend the event itself to minimize their risk of exposure.

\subsection{Case Study}
We present how the thought summary and daily plan of Isabella Rodriguez, owner of Hobbs Cafe, differed between the disease-threat and no-threat conditions during the third run on February 14th, when the Valentine’s Day party was postponed. Summaries from the first run (in which the party was held as planned) and the second run (in which the party was also postponed) are provided in Sec.~\ref{sec:app_b}.

\textit{Disease-Threat Condition}
\begin{lstlisting}
[Current status] On Tuesday, February 14th, 2023, Isabella Rodriguez is <@\textbf{feeling}@> <@\textbf{concerned}@> about the swine flu epidemic and its impact on community health. She has postponed the Valentine's Day party at Hobbs Cafe to protect her customers and friends from potential infection. Instead, she is focusing on maintaining safe daily cafe operations with enhanced hygiene practices and communicating the postponement of the event in person to her regular customers throughout the day.

[Plan for the day] Waking up and morning routine from 6 to 7 a.m., Preparing for cafe opening and personal hygiene from 7 to 8 a.m., Opening Hobbs Cafe at 8 a.m. and starting enhanced hygiene practices., <@\textbf{Communicating}@> <@\textbf{the}@> <@\textbf{postponement}@> of the Valentine's Day party to regular customers in person between 9 a.m. and 5 p.m. while working at the cafe counter., Continuing to work at the cafe counter and <@\textbf{maintain}@> <@\textbf{hygiene}@> <@\textbf{practices}@> from 5 to 8 p.m., Closing Hobbs Cafe at 8 p.m., Relaxing and having dinner from 8 to 10 p.m., Preparing for bed and unwinding from 10 to 11 p.m., Going to sleep at 11 p.m.
\end{lstlisting}

\textit{No-Threat Condition}
\begin{lstlisting}
[Current status] Isabella Rodriguez is looking forward to hosting the Valentine's Day party at Hobbs Cafe on Tuesday, February 14th, 2023, from 5 p.m. to 7 p.m. She is currently finalizing party materials and ensuring all preparations are in place. Isabella continues to invite her customers and friends in person, encouraging them to attend the celebration. She <@\textbf{feels}@> <@\textbf{excited}@> <@\textbf{and}@> <@\textbf{fulfilled}@> by the event planning and is balancing her daily responsibilities with the final touches for the party. She plans to be fully present at Hobbs Cafe during the event hours to welcome guests and create a warm, enjoyable atmosphere.

[Plan for the day] Waking up and morning routine from 6 to 7:30 a.m., Having breakfast and preparing for the day from 7:30 to 8 a.m., Opening Hobbs Cafe and standing at the counter from 8 a.m. to 3:30 p.m., Finalizing Valentine's Day <@\textbf{party}@> <@\textbf{preparation}@> at Hobbs Cafe from 3:30 to 5 p.m., <@\textbf{Hosting}@> <@\textbf{and}@> <@\textbf{being}@> <@\textbf{fully}@> <@\textbf{present}@> at the Valentine's Day party at Hobbs Cafe from 5 to 7 p.m., Continuing to stand at the counter and closing Hobbs Cafe from 7 to 8 p.m., Having dinner and relaxing from 8 to 10:30 p.m., Getting ready for bed and winding down from 10:30 to 11 p.m., Going to sleep at 11 p.m.
\end{lstlisting}

As shown above, in the disease-threat condition, Isabella expressed concern about the epidemic and prioritized community health, leading her to postpone the party. She focused on maintaining hygiene practices at the cafe and communicating the postponement to customers. For example, when Eddy Lin visited the cafe around 2:30 p.m., Isabella said, “Hi Eddy, I wanted to let you know we’ve postponed the Valentine’s Day party at Hobbs Café to keep everyone safe due to the swine flu concerns.” Eddy replied, “Thanks for letting me know, Isabella. Stay safe and let me know if there’s a new date for the party.” In contrast, in the no-threat condition, Isabella enthusiastically planned the party and invited customers.

\subsection{Validity Check: H1N1 Flu vs. Type 2 Diabetes}
In the noninfectious-disease condition, which involved a news article about type 2 diabetes, seven agents attended the party, staying for an average of 86.26 minutes. They spent an average of 98.43 minutes in third places and took an average of 598.12 steps on the day. As for social interactions, agents had a 70.0\% likelihood of initiating conversations (67.0\% with familiar agents and 75.1\% with unfamiliar agents). There were a total of 148 conversations in the town (93 with familiar and 55 with unfamiliar agents), with conversations lasting an average of 5.84 turns (5.79 with familiar agents, 6.05 with unfamiliar agents). Thirteen agents were aware of the Valentine’s Day party. On the questionnaire, they reported average scores of 3.29 for disease avoidance, 4.04 for exclusion concern, 6.41 for group affiliation, and 2.99 for independence.

Overall, the level of social engagement closely resembles that observed in the no-threat condition rather than the disease-threat (swine flu) condition. Agents expressed concern about the recent increase in hospitalizations related to type 2 diabetes and wanted to support neighbors who might be affected, but they did not avoid social interactions. Notably, after reading the article, Isabella planned to raise awareness about diabetes during the Valentine’s Day party. Thus, agents recognized that social interactions increase the risk of transmitting infectious diseases but did not associate socializing with increased risks for noninfectious conditions, which supports the validity of our experimental design.

\section{Discussion}
Across three simulation runs with generative agents, the perceived threat of swine flu led to notable decreases in sociality at both the individual and community levels. Compared to the no-threat condition, agents in the disease-threat condition showed decreased spatial engagement after reading a news article about swine flu epidemic, as seen in lower attendance and shorter stays at the Valentine’s Day party, as well as less frequent use of third places. Social interactions were similarly affected, with agents less likely to initiate conversations, participating in fewer interactions overall, and holding shorter conversations when interactions did occur. At the community level, these changes collectively resulted in slower diffusion of party-related information among agents, reflecting reduced social connectivity and movement within the town. Interviews and a case study of cafe owner Isabella Rodriguez highlighted that concerns about infection risk directly motivated these behavioral changes, as agents prioritized disease avoidance and practical responsibilities over social affiliation. Additionally, the validity check using a noninfectious disease condition (type 2 diabetes) confirmed that agents clearly distinguished between infectious and noninfectious disease threats, further strengthening confidence in our experimental design.

To our knowledge, this study is the first to employ GABM to empirically test hypotheses related to the behavioral immune system. By priming agents simply with news about the swine flu epidemic (similar to standard practices in social psychology experiments), we effectively avoided the complexity and rigidity inherent in conventional rule-based agent models. Thus, GABM presents a novel tool for social psychology research, enabling realistic simulations of both physical and social environments and allowing preliminary tests of hypotheses that may otherwise be methodologically or ethically challenging.

Nevertheless, the GABM approach has certain limitations. Current LLMs chiefly rely on data from WEIRD (Western, Educated, Industrialized, Rich, Democratic) populations, raising concerns about their generalizability and representativeness \cite{dillion2023can, henrich2010weirdest, santurkar2023whose}. Similarly, although LLMs mimic many aspects of human psychology \cite{aher2023using, hu2025generative, schramowski2022large, simmons2022moral}, 
they still lack the realism of human studies; LLMs, especially those aligned via reinforcement learning with human preferences, do not fully capture how humans respond to variations in survey questions \cite{abdurahman2024perils, tjuatja2024llms}. Moreover, LLMs remain prone to occasional hallucinations \cite{azamfirei2023large}. During our pilot testing, for example, some agents attempted to schedule meetings via Zoom despite the platform not being part of our simulated environment. Therefore, we limited interactions strictly to face-to-face communication. Continued advancements in LLM, combined with careful characterization of agent behavior by researchers, will likely mitigate these concerns in future research.

Our findings leave open numerous opportunities for researchers. Beyond disease prevalence, manipulating (perceived or actual) ecological and socioecological dimensions (e.g., population density, resource availability, walkability, residential mobility) could yield new insights into how macroenvironmental factors influence social cognition and behavior. To conduct such studies at scale, simulating larger populations of agents over extended periods would be valuable, as it could reveal long-term dynamics, including the emergence of social groups, norms, and microcultures \cite{baumard2025ecological}. Platforms like the recently developed life simulation game inZOI \cite{krafton2024inzoi}, which employs small language models to simulate around 300 characters across multiple virtual cities, may offer promising frameworks for conducting large-scale experimental investigations. In conclusion, we hope our work highlights the potential of GABM as a new methodological tool for experimental investigations in social psychology and the social sciences more broadly.

\bibliographystyle{icml2025}
\bibliography{main}

\newpage
\appendix
\onecolumn
\section{Method}\label{sec:app_a}

\subsection{Map of Smallville (Figure from \citealp{park2023generative})}
\begin{figure}[h]
    \centering
    \includegraphics[width=0.9\linewidth]{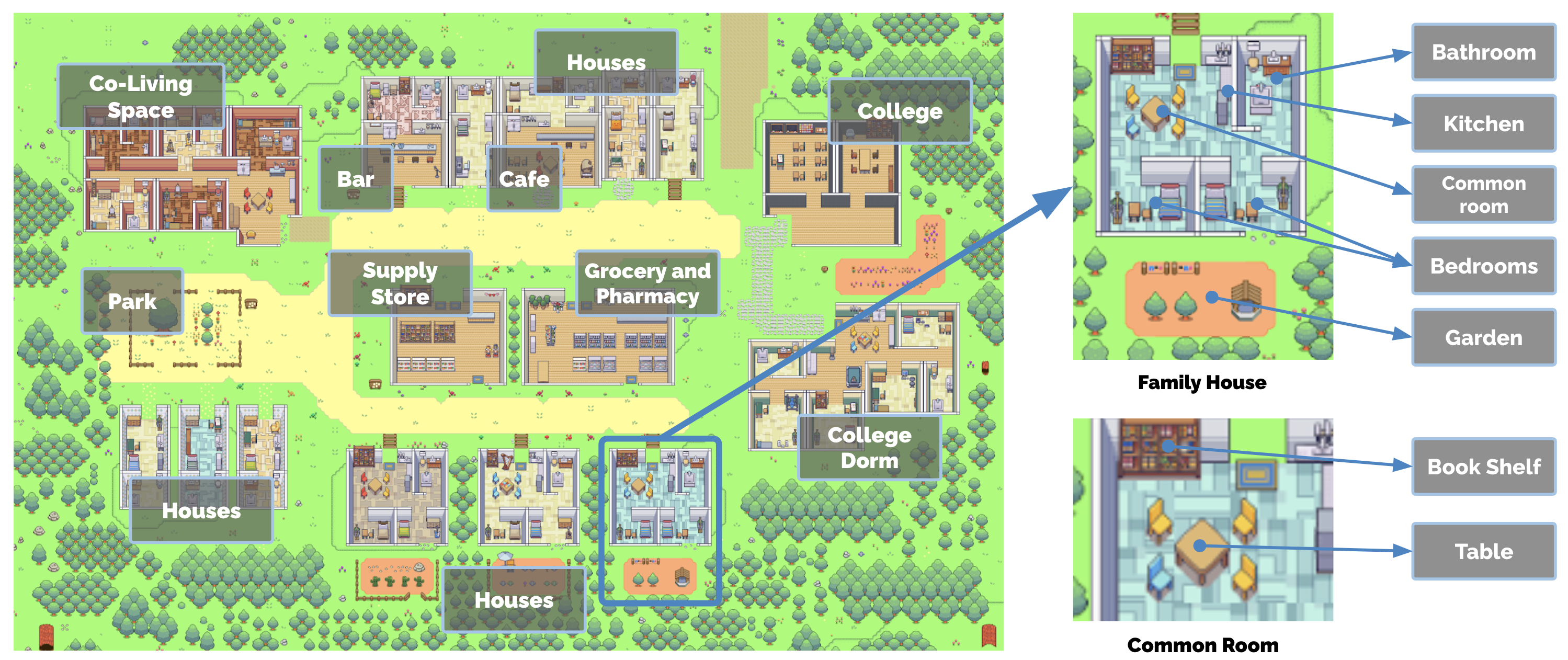}
    \label{fig:appendix_map}
\end{figure}

\subsection{Prompt for Planning the Next Day (Disease-Threat Condition)}
\begin{lstlisting}
# Role and Objective
Your goal is to update the detailed status for <@\textcolor{blue}{\textsc{character\_name}}@> for tomorrow (<@\textcolor{blue}{\textsc{tomorrow\_date}}@>) based on previous status, initial thoughts and emotions, and initial plan notes (all before reading the local news), as well as the new context provided by the local news. 

# Instructions
Given the provided information about <@\textcolor{blue}{\textsc{character\_name}}@> and the new context from local news, carefully think step-by-step about how these together might shape <@\textcolor{blue}{\textsc{character\_name}}@>'s thoughts, emotions, behaviors, social interactions, and daily activities. Then, clearly update <@\textcolor{blue}{\textsc{character\_name}}@>'s thoughts and emotions, plan notes, and detailed status. You have the flexibility to add, remove, or adjust scheduled activities in the plan note. Write in the third person, talking about <@\textcolor{blue}{\textsc{character\_name}}@>. For any scheduling-related information, clearly specify all essential details, including the exact date, time, and location. Do not assume actions planned for <@\textcolor{blue}{\textsc{tomorrow\_date}}@> have already been completed or that communication has already occurred unless explicitly stated in the context. Remember, there is no remote communication (phone, internet, etc.); communication requires face-to-face interaction.

# Output Format
Respond ONLY with a JSON object structured as follows:
{
  "reasoning": "<brief_reason>",
  "updated_thought_and_emotion_note": "<string>",
  "updated_plan_note": "<string>",
  "new_status": "<string>"
}

# Context
<@\textcolor{blue}{\textsc{character\_name}}@>'s status from <@\textcolor{blue}{\textsc{today\_date}}@>:
<@\textcolor{blue}{\textsc{character\_current\_status}}@>

<@\textcolor{blue}{\textsc{character\_name}}@>'s initial thoughts and emotions (before reading the news):
<@\textcolor{blue}{\textsc{character\_thought\_note}}@>

<@\textcolor{blue}{\textsc{character\_name}}@>'s initial plan notes for <@\textcolor{blue}{\textsc{tomorrow\_date}}@> (before reading the news):
<@\textcolor{blue}{\textsc{character\_plan\_note}}@>

<@\textcolor{blue}{\textsc{news\_article}}@>

Reading the news reminded <@\textcolor{blue}{\textsc{character\_name}}@> that a few neighbors had seemed unwell lately.

# Final Instructions
Given the information above, carefully think step-by-step and write in the third person:
1. Clearly explain how the local news context might affect <@\textcolor{blue}{\textsc{character\_name}}@>'s thoughts, emotions, behaviors, social interactions, and daily activities.
2. Write <@\textcolor{blue}{\textsc{character\_name}}@>'s updated thoughts and emotions after reading the local news.
3. Write <@\textcolor{blue}{\textsc{character\_name}}@>'s updated plan notes after reading the local news. You have the flexibility to add, remove, or adjust scheduled activities in the plan note.
4. Clearly state <@\textcolor{blue}{\textsc{character\_name}}@>'s detailed status for <@\textcolor{blue}{\textsc{tomorrow\_date}}@> based on the updated thoughts and emotions and the updated plan notes. 
For any scheduling-related information, clearly specify all essential details, including the exact date, time, and location. The status should only include information grounded in the context provided and reflect the state *before* any actions planned for <@\textcolor{blue}{\textsc{tomorrow\_date}}@> are taken. Remember, communication requires physical presence.
\end{lstlisting}

\subsection{Prompt for Planning the Next Day (No-Threat Condition)}
\begin{lstlisting}

# Role and Objective
Your goal is to update the detailed status for <@\textcolor{blue}{\textsc{character\_name}}@> for tomorrow (<@\textcolor{blue}{\textsc{tomorrow\_date}}@>) based on previous status, thoughts and emotions, and plan notes.

# Instructions
Given <@\textcolor{blue}{\textsc{character\_name}}@>'s previous status, thoughts and emotions, and plan notes, carefully think step-by-step about how these together shape the detailed status for <@\textcolor{blue}{\textsc{tomorrow\_date}}@>. You have the flexibility to add, remove, or adjust scheduled activities in the plan note. Write in the third person, talking about <@\textcolor{blue}{\textsc{character\_name}}@>. For any scheduling-related information, clearly specify all essential details, including the exact date, time, and location. Do not assume actions planned for <@\textcolor{blue}{\textsc{tomorrow\_date}}@> have already been completed or that communication has already occurred unless explicitly stated in the context. Remember, there is no remote communication (phone, internet, etc.); communication requires face-to-face interaction.

# Output Format
Respond ONLY with a JSON object structured as follows:
{
  "reasoning": "<brief_reason>",
  "new_status": "<string>"
}

# Context
<@\textcolor{blue}{\textsc{character\_name}}@>'s status from <@\textcolor{blue}{\textsc{today\_date}}@>:
<@\textcolor{blue}{\textsc{character\_current\_status}}@>

<@\textcolor{blue}{\textsc{character\_name}}@>'s thoughts and emotions:
<@\textcolor{blue}{\textsc{character\_thought\_note}}@>

<@\textcolor{blue}{\textsc{character\_name}}@>'s plan note for <@\textcolor{blue}{\textsc{tomorrow\_date}}@>:
<@\textcolor{blue}{\textsc{character\_plan\_note}}@>

# Final Instructions
Given the information above, carefully think step-by-step, and then clearly state <@\textcolor{blue}{\textsc{character\_name}}@>'s detailed status for <@\textcolor{blue}{\textsc{tomorrow\_date}}@>. Write in the third person. You have the flexibility to add, remove, or adjust scheduled activities in the plan note. For any scheduling-related information, clearly specify all essential details, including the exact date, time, and location. The status should only include information grounded in the context provided and reflect the state *before* any actions planned for <@\textcolor{blue}{\textsc{tomorrow\_date}}@> are taken. Remember, communication requires physical presence.
\end{lstlisting}

\subsection{News Article in Noninfectious-Disease Condition}
\begin{quoting}
\emph{The latest study of Type 2 diabetes complications suggested that one quarter of people in our country sick enough to be hospitalized with diabetes last year also wound up needing intensive care; of these, roughly 7 percent of them died. These rates are higher than with many other chronic illnesses, and experts note that what is striking and unusual is that people who previously had their condition under control accounted for nearly half of the hospitalized cases. In total, health authorities claim that over 2 million people across the country could be hospitalized because of diabetes-related complications. They also noted that individuals might be experiencing complications even while asymptomatic, increasing the risk of unnoticed progression in our community.}
\end{quoting}

\newpage
\subsection{Relationship Mapping}
Below is a table showing the relationship mapping among the 25 agents on the morning of February 13.
\begin{center}
\begin{table}[h]
\centering
  \caption{\textit{Relationship Mapping}}
  \label{tab:relationship_mapping}
\tiny %
\setlength{\tabcolsep}{1pt} %
\renewcommand{\arraystretch}{1} %

\begin{tabular}{|p{2.3cm}|*{25}{c|}}
\hline
\textbf{Person} &
\rotatebox{90}{Abigail Chen} &
\rotatebox{90}{Adam Smith} &
\rotatebox{90}{Arthur Burton} &
\rotatebox{90}{Ayesha Khan} &
\rotatebox{90}{Carlos Gomez} &
\rotatebox{90}{Carmen Ortiz} &
\rotatebox{90}{Eddy Lin} &
\rotatebox{90}{Francisco Lopez} &
\rotatebox{90}{Giorgio Rossi} &
\rotatebox{90}{Hailey Johnson} &
\rotatebox{90}{Isabella Rodriguez} &
\rotatebox{90}{Jane Moreno} &
\rotatebox{90}{Jennifer Moore} &
\rotatebox{90}{John Lin} &
\rotatebox{90}{Klaus Mueller} &
\rotatebox{90}{Latoya Williams} &
\rotatebox{90}{Maria Lopez} &
\rotatebox{90}{Mei Lin} &
\rotatebox{90}{Rajiv Patel} &
\rotatebox{90}{Ryan Park} &
\rotatebox{90}{Sam Moore} &
\rotatebox{90}{Tamara Taylor} &
\rotatebox{90}{Tom Moreno} &
\rotatebox{90}{Wolfgang Schulz} &
\rotatebox{90}{Yuriko Yamamoto} \\
\hline
\centeredrowheader{Abigail Chen} & \squarecell{} & \squarecell{} & \squarecell{} & \squarecell{} & \squarecell{} & \squarecell{} & \squarecell{} & \squarecell{\fancymark} & \squarecell{} & \squarecell{\fancymark} & \squarecell{} & \squarecell{} & \squarecell{} & \squarecell{} & \squarecell{} & \squarecell{\fancymark} & \squarecell{} & \squarecell{} & \squarecell{\fancymark} & \squarecell{} & \squarecell{} & \squarecell{} & \squarecell{} & \squarecell{} & \squarecell{} \\ \hline
\centeredrowheader{Adam Smith} & \squarecell{} & \squarecell{} & \squarecell{\fancymark} & \squarecell{} & \squarecell{} & \squarecell{\fancymark} & \squarecell{} & \squarecell{} & \squarecell{} & \squarecell{} & \squarecell{} & \squarecell{\fancymark} & \squarecell{\fancymark} & \squarecell{} & \squarecell{} & \squarecell{} & \squarecell{} & \squarecell{} & \squarecell{} & \squarecell{} & \squarecell{\fancymark} & \squarecell{\fancymark} & \squarecell{\fancymark} & \squarecell{} & \squarecell{\fancymark} \\ \hline
\centeredrowheader{Arthur Burton} & \squarecell{} & \squarecell{\fancymark} & \squarecell{} & \squarecell{} & \squarecell{\fancymark} & \squarecell{\fancymark} & \squarecell{} & \squarecell{} & \squarecell{} & \squarecell{} & \squarecell{\fancymark} & \squarecell{} & \squarecell{\fancymark} & \squarecell{} & \squarecell{} & \squarecell{\fancymark} & \squarecell{} & \squarecell{} & \squarecell{\fancymark} & \squarecell{\fancymark} & \squarecell{\fancymark} & \squarecell{} & \squarecell{\fancymark} & \squarecell{} & \squarecell{\fancymark} \\ \hline
\centeredrowheader{Ayesha Khan} & \squarecell{} & \squarecell{} & \squarecell{} & \squarecell{} & \squarecell{} & \squarecell{} & \squarecell{\fancymark} & \squarecell{} & \squarecell{} & \squarecell{} & \squarecell{} & \squarecell{} & \squarecell{} & \squarecell{} & \squarecell{\fancymark} & \squarecell{} & \squarecell{\fancymark} & \squarecell{\fancymark} & \squarecell{} & \squarecell{} & \squarecell{} & \squarecell{} & \squarecell{} & \squarecell{\fancymark} & \squarecell{} \\ \hline
\centeredrowheader{Carlos Gomez} & \squarecell{} & \squarecell{} & \squarecell{\fancymark} & \squarecell{} & \squarecell{} & \squarecell{} & \squarecell{} & \squarecell{} & \squarecell{\fancymark} & \squarecell{} & \squarecell{\fancymark} & \squarecell{} & \squarecell{} & \squarecell{} & \squarecell{} & \squarecell{} & \squarecell{} & \squarecell{} & \squarecell{} & \squarecell{\fancymark} & \squarecell{} & \squarecell{} & \squarecell{} & \squarecell{} & \squarecell{} \\ \hline
\centeredrowheader{Carmen Ortiz} & \squarecell{} & \squarecell{\fancymark} & \squarecell{\fancymark} & \squarecell{} & \squarecell{} & \squarecell{} & \squarecell{\fancymark} & \squarecell{} & \squarecell{} & \squarecell{} & \squarecell{\fancymark} & \squarecell{\fancymark} & \squarecell{\fancymark} & \squarecell{\fancymark} & \squarecell{} & \squarecell{} & \squarecell{} & \squarecell{\fancymark} & \squarecell{} & \squarecell{} & \squarecell{\fancymark} & \squarecell{\fancymark} & \squarecell{\fancymark} & \squarecell{} & \squarecell{\fancymark} \\ \hline
\centeredrowheader{Eddy Lin} & \squarecell{} & \squarecell{} & \squarecell{} & \squarecell{\fancymark} & \squarecell{} & \squarecell{\fancymark} & \squarecell{} & \squarecell{} & \squarecell{} & \squarecell{} & \squarecell{} & \squarecell{\fancymark} & \squarecell{\fancymark} & \squarecell{\fancymark} & \squarecell{\fancymark} & \squarecell{} & \squarecell{\fancymark} & \squarecell{\fancymark} & \squarecell{} & \squarecell{} & \squarecell{\fancymark} & \squarecell{\fancymark} & \squarecell{\fancymark} & \squarecell{\fancymark} & \squarecell{\fancymark} \\ \hline
\centeredrowheader{Francisco Lopez} & \squarecell{\fancymark} & \squarecell{} & \squarecell{} & \squarecell{} & \squarecell{} & \squarecell{} & \squarecell{} & \squarecell{} & \squarecell{} & \squarecell{\fancymark} & \squarecell{} & \squarecell{} & \squarecell{} & \squarecell{} & \squarecell{} & \squarecell{\fancymark} & \squarecell{} & \squarecell{} & \squarecell{\fancymark} & \squarecell{} & \squarecell{} & \squarecell{} & \squarecell{} & \squarecell{} & \squarecell{} \\ \hline
\centeredrowheader{Giorgio Rossi} & \squarecell{} & \squarecell{} & \squarecell{} & \squarecell{} & \squarecell{\fancymark} & \squarecell{} & \squarecell{} & \squarecell{} & \squarecell{} & \squarecell{} & \squarecell{\fancymark} & \squarecell{} & \squarecell{} & \squarecell{} & \squarecell{} & \squarecell{} & \squarecell{} & \squarecell{} & \squarecell{} & \squarecell{\fancymark} & \squarecell{} & \squarecell{} & \squarecell{} & \squarecell{} & \squarecell{} \\ \hline
\centeredrowheader{Hailey Johnson} & \squarecell{\fancymark} & \squarecell{} & \squarecell{} & \squarecell{} & \squarecell{} & \squarecell{} & \squarecell{} & \squarecell{\fancymark} & \squarecell{} & \squarecell{} & \squarecell{\fancymark} & \squarecell{} & \squarecell{} & \squarecell{} & \squarecell{} & \squarecell{\fancymark} & \squarecell{} & \squarecell{} & \squarecell{\fancymark} & \squarecell{} & \squarecell{} & \squarecell{} & \squarecell{} & \squarecell{} & \squarecell{} \\ \hline
\centeredrowheader{Isabella Rodriguez} & \squarecell{} & \squarecell{} & \squarecell{\fancymark} & \squarecell{} & \squarecell{\fancymark} & \squarecell{\fancymark} & \squarecell{} & \squarecell{} & \squarecell{\fancymark} & \squarecell{\fancymark} & \squarecell{} & \squarecell{} & \squarecell{} & \squarecell{} & \squarecell{} & \squarecell{} & \squarecell{\fancymark} & \squarecell{} & \squarecell{} & \squarecell{\fancymark} & \squarecell{} & \squarecell{\fancymark} & \squarecell{} & \squarecell{\fancymark} & \squarecell{} \\ \hline
\centeredrowheader{Jane Moreno} & \squarecell{} & \squarecell{\fancymark} & \squarecell{} & \squarecell{} & \squarecell{} & \squarecell{\fancymark} & \squarecell{\fancymark} & \squarecell{} & \squarecell{} & \squarecell{} & \squarecell{} & \squarecell{} & \squarecell{\fancymark} & \squarecell{\fancymark} & \squarecell{} & \squarecell{} & \squarecell{} & \squarecell{\fancymark} & \squarecell{} & \squarecell{} & \squarecell{\fancymark} & \squarecell{\fancymark} & \squarecell{\fancymark} & \squarecell{} & \squarecell{\fancymark} \\ \hline
\centeredrowheader{Jennifer Moore} & \squarecell{} & \squarecell{\fancymark} & \squarecell{\fancymark} & \squarecell{} & \squarecell{} & \squarecell{\fancymark} & \squarecell{\fancymark} & \squarecell{} & \squarecell{} & \squarecell{} & \squarecell{} & \squarecell{\fancymark} & \squarecell{} & \squarecell{\fancymark} & \squarecell{} & \squarecell{} & \squarecell{} & \squarecell{\fancymark} & \squarecell{} & \squarecell{} & \squarecell{\fancymark} & \squarecell{} & \squarecell{\fancymark} & \squarecell{} & \squarecell{\fancymark} \\ \hline
\centeredrowheader{John Lin} & \squarecell{} & \squarecell{} & \squarecell{} & \squarecell{} & \squarecell{} & \squarecell{\fancymark} & \squarecell{\fancymark} & \squarecell{} & \squarecell{} & \squarecell{} & \squarecell{} & \squarecell{\fancymark} & \squarecell{\fancymark} & \squarecell{} & \squarecell{} & \squarecell{} & \squarecell{} & \squarecell{\fancymark} & \squarecell{} & \squarecell{} & \squarecell{\fancymark} & \squarecell{\fancymark} & \squarecell{\fancymark} & \squarecell{} & \squarecell{\fancymark} \\ \hline
\centeredrowheader{Klaus Mueller} & \squarecell{} & \squarecell{} & \squarecell{} & \squarecell{\fancymark} & \squarecell{} & \squarecell{} & \squarecell{\fancymark} & \squarecell{} & \squarecell{} & \squarecell{} & \squarecell{} & \squarecell{} & \squarecell{} & \squarecell{} & \squarecell{} & \squarecell{} & \squarecell{\fancymark} & \squarecell{\fancymark} & \squarecell{} & \squarecell{} & \squarecell{} & \squarecell{} & \squarecell{} & \squarecell{\fancymark} & \squarecell{} \\ \hline
\centeredrowheader{Latoya Williams} & \squarecell{\fancymark} & \squarecell{} & \squarecell{\fancymark} & \squarecell{} & \squarecell{} & \squarecell{} & \squarecell{} & \squarecell{\fancymark} & \squarecell{} & \squarecell{\fancymark} & \squarecell{} & \squarecell{} & \squarecell{} & \squarecell{} & \squarecell{} & \squarecell{} & \squarecell{} & \squarecell{} & \squarecell{\fancymark} & \squarecell{} & \squarecell{} & \squarecell{} & \squarecell{} & \squarecell{} & \squarecell{} \\ \hline
\centeredrowheader{Maria Lopez} & \squarecell{} & \squarecell{} & \squarecell{} & \squarecell{\fancymark} & \squarecell{} & \squarecell{} & \squarecell{\fancymark} & \squarecell{} & \squarecell{} & \squarecell{} & \squarecell{\fancymark} & \squarecell{} & \squarecell{} & \squarecell{} & \squarecell{\fancymark} & \squarecell{} & \squarecell{} & \squarecell{\fancymark} & \squarecell{} & \squarecell{} & \squarecell{} & \squarecell{} & \squarecell{} & \squarecell{\fancymark} & \squarecell{} \\ \hline
\centeredrowheader{Mei Lin} & \squarecell{} & \squarecell{} & \squarecell{} & \squarecell{\fancymark} & \squarecell{} & \squarecell{\fancymark} & \squarecell{\fancymark} & \squarecell{} & \squarecell{} & \squarecell{} & \squarecell{} & \squarecell{\fancymark} & \squarecell{\fancymark} & \squarecell{\fancymark} & \squarecell{\fancymark} & \squarecell{} & \squarecell{\fancymark} & \squarecell{} & \squarecell{} & \squarecell{} & \squarecell{\fancymark} & \squarecell{\fancymark} & \squarecell{\fancymark} & \squarecell{\fancymark} & \squarecell{\fancymark} \\ \hline
\centeredrowheader{Rajiv Patel} & \squarecell{\fancymark} & \squarecell{} & \squarecell{\fancymark} & \squarecell{} & \squarecell{} & \squarecell{} & \squarecell{} & \squarecell{\fancymark} & \squarecell{} & \squarecell{\fancymark} & \squarecell{} & \squarecell{} & \squarecell{} & \squarecell{} & \squarecell{} & \squarecell{\fancymark} & \squarecell{} & \squarecell{} & \squarecell{} & \squarecell{} & \squarecell{} & \squarecell{} & \squarecell{} & \squarecell{} & \squarecell{} \\ \hline
\centeredrowheader{Ryan Park} & \squarecell{} & \squarecell{} & \squarecell{\fancymark} & \squarecell{} & \squarecell{\fancymark} & \squarecell{} & \squarecell{} & \squarecell{} & \squarecell{\fancymark} & \squarecell{} & \squarecell{\fancymark} & \squarecell{} & \squarecell{} & \squarecell{} & \squarecell{} & \squarecell{} & \squarecell{} & \squarecell{} & \squarecell{} & \squarecell{} & \squarecell{} & \squarecell{} & \squarecell{} & \squarecell{} & \squarecell{} \\ \hline
\centeredrowheader{Sam Moore} & \squarecell{} & \squarecell{\fancymark} & \squarecell{\fancymark} & \squarecell{} & \squarecell{} & \squarecell{\fancymark} & \squarecell{\fancymark} & \squarecell{} & \squarecell{} & \squarecell{} & \squarecell{} & \squarecell{\fancymark} & \squarecell{\fancymark} & \squarecell{\fancymark} & \squarecell{} & \squarecell{} & \squarecell{} & \squarecell{\fancymark} & \squarecell{} & \squarecell{} & \squarecell{} & \squarecell{} & \squarecell{\fancymark} & \squarecell{} & \squarecell{\fancymark} \\ \hline
\centeredrowheader{Tamara Taylor} & \squarecell{} & \squarecell{\fancymark} & \squarecell{} & \squarecell{} & \squarecell{} & \squarecell{\fancymark} & \squarecell{\fancymark} & \squarecell{} & \squarecell{} & \squarecell{} & \squarecell{\fancymark} & \squarecell{\fancymark} & \squarecell{} & \squarecell{\fancymark} & \squarecell{} & \squarecell{} & \squarecell{} & \squarecell{\fancymark} & \squarecell{} & \squarecell{} & \squarecell{} & \squarecell{} & \squarecell{\fancymark} & \squarecell{} & \squarecell{} \\ \hline
\centeredrowheader{Tom Moreno} & \squarecell{} & \squarecell{\fancymark} & \squarecell{\fancymark} & \squarecell{} & \squarecell{} & \squarecell{\fancymark} & \squarecell{\fancymark} & \squarecell{} & \squarecell{} & \squarecell{} & \squarecell{} & \squarecell{\fancymark} & \squarecell{\fancymark} & \squarecell{\fancymark} & \squarecell{} & \squarecell{} & \squarecell{} & \squarecell{\fancymark} & \squarecell{} & \squarecell{} & \squarecell{\fancymark} & \squarecell{} & \squarecell{} & \squarecell{} & \squarecell{\fancymark} \\ \hline
\centeredrowheader{Wolfgang Schulz} & \squarecell{} & \squarecell{} & \squarecell{} & \squarecell{\fancymark} & \squarecell{} & \squarecell{} & \squarecell{\fancymark} & \squarecell{} & \squarecell{} & \squarecell{} & \squarecell{\fancymark} & \squarecell{} & \squarecell{} & \squarecell{} & \squarecell{\fancymark} & \squarecell{} & \squarecell{\fancymark} & \squarecell{\fancymark} & \squarecell{} & \squarecell{} & \squarecell{} & \squarecell{} & \squarecell{} & \squarecell{} & \squarecell{} \\ \hline
\centeredrowheader{Yuriko Yamamoto} & \squarecell{} & \squarecell{\fancymark} & \squarecell{\fancymark} & \squarecell{} & \squarecell{} & \squarecell{\fancymark} & \squarecell{\fancymark} & \squarecell{} & \squarecell{} & \squarecell{} & \squarecell{} & \squarecell{\fancymark} & \squarecell{\fancymark} & \squarecell{\fancymark} & \squarecell{} & \squarecell{} & \squarecell{} & \squarecell{\fancymark} & \squarecell{} & \squarecell{} & \squarecell{\fancymark} & \squarecell{} & \squarecell{\fancymark} & \squarecell{} & \squarecell{} \\ \hline
\end{tabular}
\end{table}
\end{center}

\newpage
\section{Detailed Results}\label{sec:app_b}
\subsection{Valentine’s Day Party Attendance List (Excluding Isabella Rodriguez)}
\textit{Disease-Threat Condition}
\begin{itemize}
    \item Run 1: Carmen Ortiz
    \item Run 2 (no party): Latoya Williams, Klaus Mueller (independently of the party)
    \item Run 3 (no party): Sam Moore (attended the party but was informed of its postponement by Isabella)
\end{itemize}
\textit{No-Threat Condition}
\begin{itemize}
    \item Run 1: Ayesha Khan, Carmen Ortiz, Eddy Lin, Giorgio Rossi, Klaus Mueller, Maria Lopez, Ryan Park, Sam Moore
    \item Run 2: Ayesha Khan, Giorgio Rossi, Klaus Mueller, Latoya Williams, Maria Lopez, Sam Moore
    \item Run 3: Ayesha Khan, Carlos Gomez, Eddy Lin, Giorgio Rossi, Klaus Mueller, Maria Lopez, Sam Moore, Wolfgang Schulz
\end{itemize}
\textit{Noninfectious-Disease Condition}
\begin{itemize}
    \item Run 1: Ayesha Khan, Carmen Ortiz, Eddy Lin, Giorgio Rossi, Maria Lopez, Ryan Park, Sam Moore
\end{itemize}

\vspace{0.1 in}

\subsection{Detailed Findings from Simulation Runs}
\begin{table*}[h]
  \centering
  \scriptsize
  \setlength{\tabcolsep}{6pt}      %
  \renewcommand{\arraystretch}{1.2} %

  \caption{\textit{Sociality on February~14}}
  \label{tab:metrics-feb14}

  \begin{tabular}{lcccccccc}
    \toprule
    & \multicolumn{4}{c}{\textbf{Disease-threat condition}}
    & \multicolumn{4}{c}{\textbf{No-threat condition}} \\
    \cmidrule(lr){2-5}\cmidrule(lr){6-9}
    Measures
      & Run~1 & Run~2 & Run~3 & \textit{M}
      & Run~1 & Run~2 & Run~3 & \textit{M} \\
    \midrule
    \addlinespace
    \multicolumn{9}{l}{\textbf{Spatial behavior}} \\
    \addlinespace
    \makecell[l]{Party attendance\\(number of agents)}
      & 1   & 2   & 1   & 1.33
      & 8   & 6   & 8   & 7.33 \\
    \addlinespace
    \makecell[l]{Party stay duration\\(in minutes)}
      & 103.17 & 53.58 & 26.00 & 60.92
      & 81.00  & 77.44 & 81.83 & 80.09 \\
    \addlinespace
    \makecell[l]{Third place time\\(in minutes)}
      & 30.52  & 44.43 & 15.39 & 30.11
      & 128.69 & 93.73 & 115.53& 112.65\\
    \addlinespace
    \makecell[l]{Physical movement\\(number of steps)}
      & 400.88 & 490.60 & 420.28 & 437.25
      & 569.36 & 546.60 & 572.12& 562.69\\
    \addlinespace
    \multicolumn{9}{l}{\textbf{Social interaction}} \\
    \addlinespace
    \makecell[l]{Conversation probability (freq)\\All (familiar, unfamiliar)}
      & \makecell[c]{49.6\\(54.6,\,20.2)} & \makecell[c]{44.7\\(52.3,\,35.8)}
      & \makecell[c]{55.3\\(68.1,\,23.0)} & \makecell[c]{49.9\\(58.3,\,26.3)}
      & \makecell[c]{61.4\\(64.3,\,53.1)} & \makecell[c]{61.3\\(74.3,\,40.3)}
      & \makecell[c]{64.0\\(70.8,\,55.2)} & \makecell[c]{62.2\\(69.8,\,49.5)} \\
    \addlinespace
    \makecell[l]{Conversation probability (p)\\All (familiar, unfamiliar)}
      & \makecell[c]{51.7\\(57.3,\,27.4)} & \makecell[c]{46.9\\(55.1,\,28.0)}
      & \makecell[c]{52.0\\(61.0,\,34.7)} & \makecell[c]{50.2\\(57.8,\,30.0)}
      & \makecell[c]{63.3\\(68.2,\,51.0)} & \makecell[c]{63.3\\(70.6,\,47.7)}
      & \makecell[c]{62.4\\(67.8,\,53.0)} & \makecell[c]{63.0\\(68.9,\,50.6)} \\
    \addlinespace
    \makecell[l]{Conversation frequency\\All (familiar, unfamiliar)}
      & \makecell[c]{75\\(64,\,11)}    & \makecell[c]{83\\(67,\,16)}
      & \makecell[c]{63\\(54,\,9)}    & \makecell[c]{73.67\\(61.67,\,12.00)}
      & \makecell[c]{153\\(104,\,49)} & \makecell[c]{125\\(82,\,43)}
      & \makecell[c]{150\\(92,\,58)}  & \makecell[c]{142.67\\(92.67,\,50.00)} \\
    \addlinespace
    \makecell[l]{Conversation length\\All (familiar, unfamiliar)}
      & \makecell[c]{5.59\\(5.48, 5.80)} & \makecell[c]{5.34\\(5.41, 5.04)}
      & \makecell[c]{5.61\\(5.94, 5.10)} & \makecell[c]{5.51\\(5.61, 5.31)}
      & \makecell[c]{5.85\\(5.91, 5.36)}   & \makecell[c]{5.51\\(5.61, 5.50)}
      & \makecell[c]{5.67\\(5.77, 5.14)} & \makecell[c]{5.68\\(5.76, 5.33)} \\
    \addlinespace
    \makecell[l]{Party diffusion\\(number of agents)}
      & 13 & 8 & 11 & 10.67
      & 13 & 12 & 13 & 12.67 \\
    \bottomrule
  \end{tabular}
\end{table*}

\vspace{0.5in}

\begin{table*}[h]
  \centering
  \scriptsize
  \setlength{\tabcolsep}{6pt}      %
  \renewcommand{\arraystretch}{1.2} %
  \caption{\textit{Sociality on February~13}}
  \label{tab:metrics-feb13}
  \begin{tabular}{lcccc}
    \toprule
    & \multicolumn{4}{c}{February~13 (for both conditions)} \\
    \cmidrule(lr){2-5}
    Measures
      & Run~1 & Run~2 & Run~3 & \textit{M} \\
    \midrule
    \addlinespace
    \multicolumn{5}{l}{\textbf{Spatial behavior}} \\
    \addlinespace
    \makecell[l]{Third place time\\(in minutes)}
      & 91.42  & 71.60  & 89.59  & 84.20  \\
    \addlinespace
    \makecell[l]{Physical movement\\(number of steps)}
      & 488.32 & 532.92 & 520.56 & 513.93 \\
    \addlinespace
    \multicolumn{5}{l}{\textbf{Social interaction}} \\
    \addlinespace
    \makecell[l]{Conversation probability (freq)\\All (familiar, unfamiliar)}
      & \makecell[c]{64.1\\(73.5,\,40.8)}
      & \makecell[c]{65.5\\(67.7,\,57.2)}
      & \makecell[c]{57.0\\(57.6,\,59.8)}
      & \makecell[c]{62.2\\(66.3,\,52.6)} \\
    \addlinespace
    \makecell[l]{Conversation probability (p)\\All (familiar, unfamiliar)}
      & \makecell[c]{58.8\\(70.2,\,37.9)}
      & \makecell[c]{62.7\\(67.8,\,51.3)}
      & \makecell[c]{64.6\\(65.8,\,49.7)}
      & \makecell[c]{62.0\\(67.9,\,46.3)} \\
    \addlinespace
    \makecell[l]{Conversation frequency\\All (familiar, unfamiliar)}
      & \makecell[c]{106\\(78,\,28)}
      & \makecell[c]{117\\(84,\,33)}
      & \makecell[c]{108\\(79,\,29)}
      & \makecell[c]{110.33\\(80.33,\,30.00)} \\
    \addlinespace
    \makecell[l]{Conversation length\\All (familiar, unfamiliar)}
      & \makecell[c]{5.92\\(6.01, 5.50)}
      & \makecell[c]{5.45\\(5.27, 5.32)}
      & \makecell[c]{5.74\\(5.57, 5.30)}
      & \makecell[c]{5.70\\(5.62, 5.37)} \\
    \bottomrule
  \end{tabular}
\end{table*}

\subsection{Valentine's Day Party Screenshots of Second and Third Runs}

\textit{Run 2 (Disease-Threat Condition)}
\begin{figure}[H]
    \centering
    \includegraphics[width=0.6\textwidth]{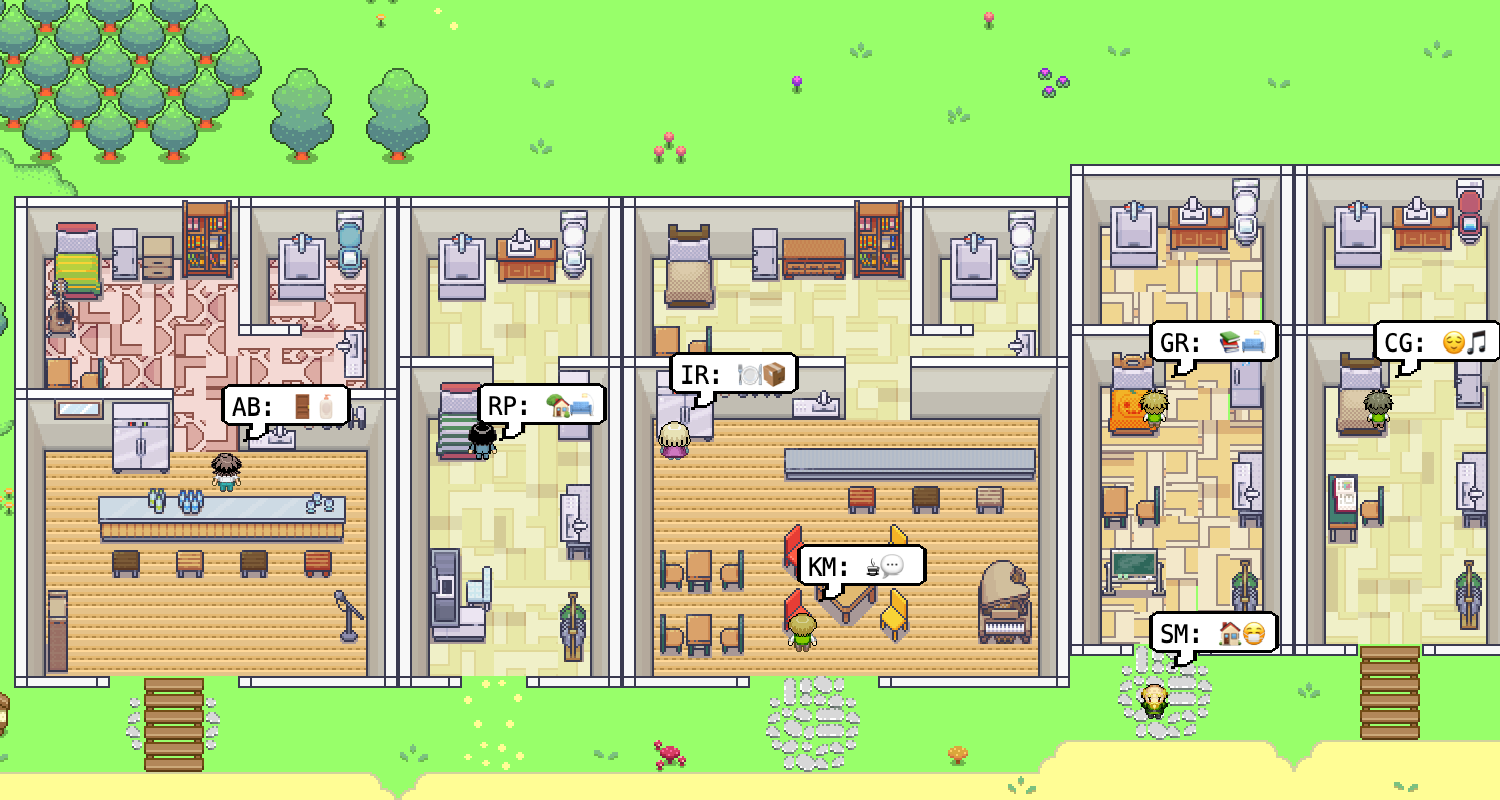}
    \label{fig:R2_true}
\end{figure}

\textit{Run 3 (Disease-Threat Condition)}
\begin{figure}[H]
    \centering
    \includegraphics[width=0.6\textwidth]{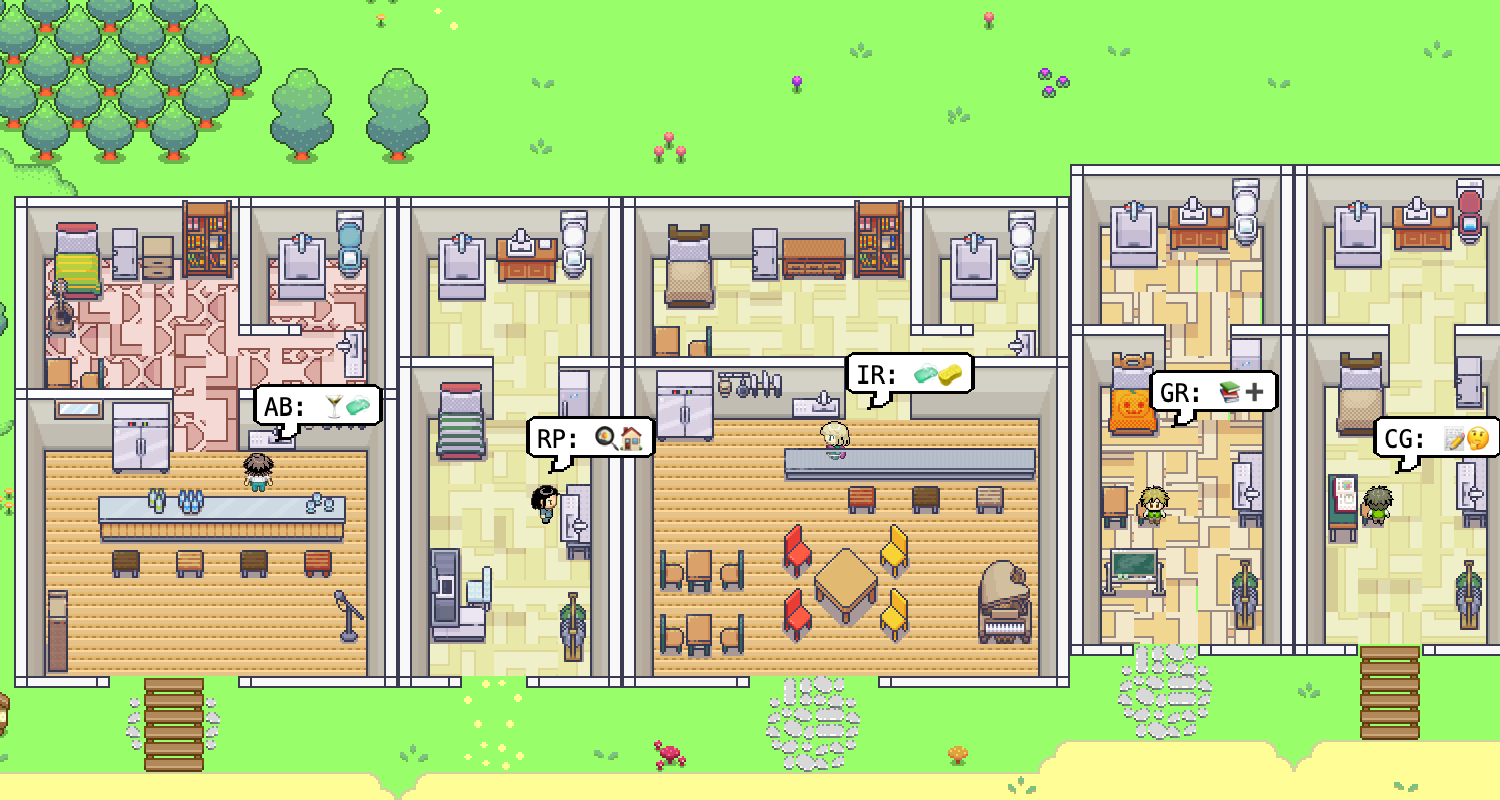}
    \label{fig:R3_true}
\end{figure}

\textit{Run 2 (No-Threat Condition)}
\begin{figure}[H]
    \centering
    \includegraphics[width=0.6\textwidth]{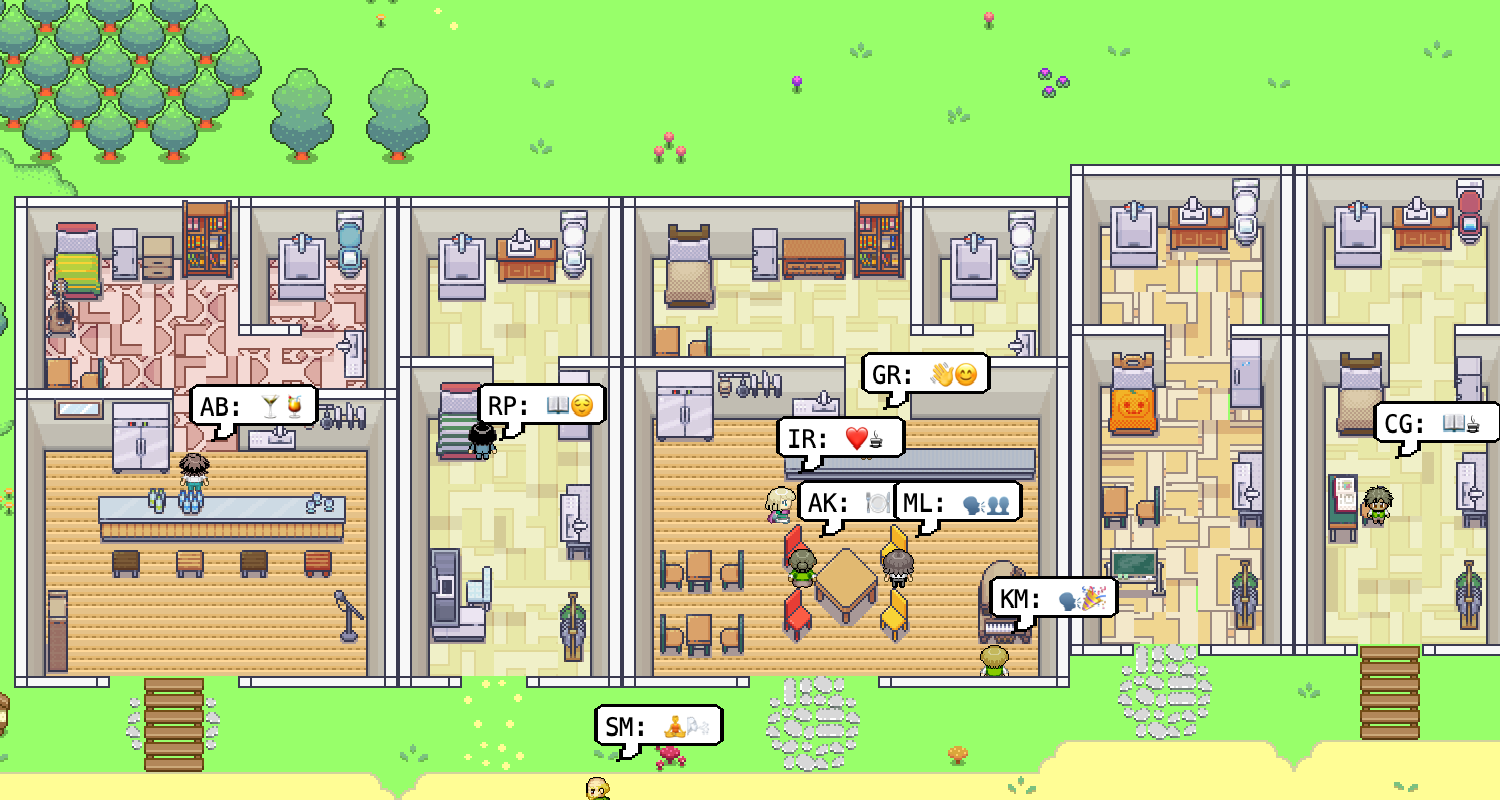}
    \label{fig:R2_false}
\end{figure}

\textit{Run 3 (No-Threat Condition)}
\begin{figure}[H]
    \centering
    \includegraphics[width=0.6\textwidth]{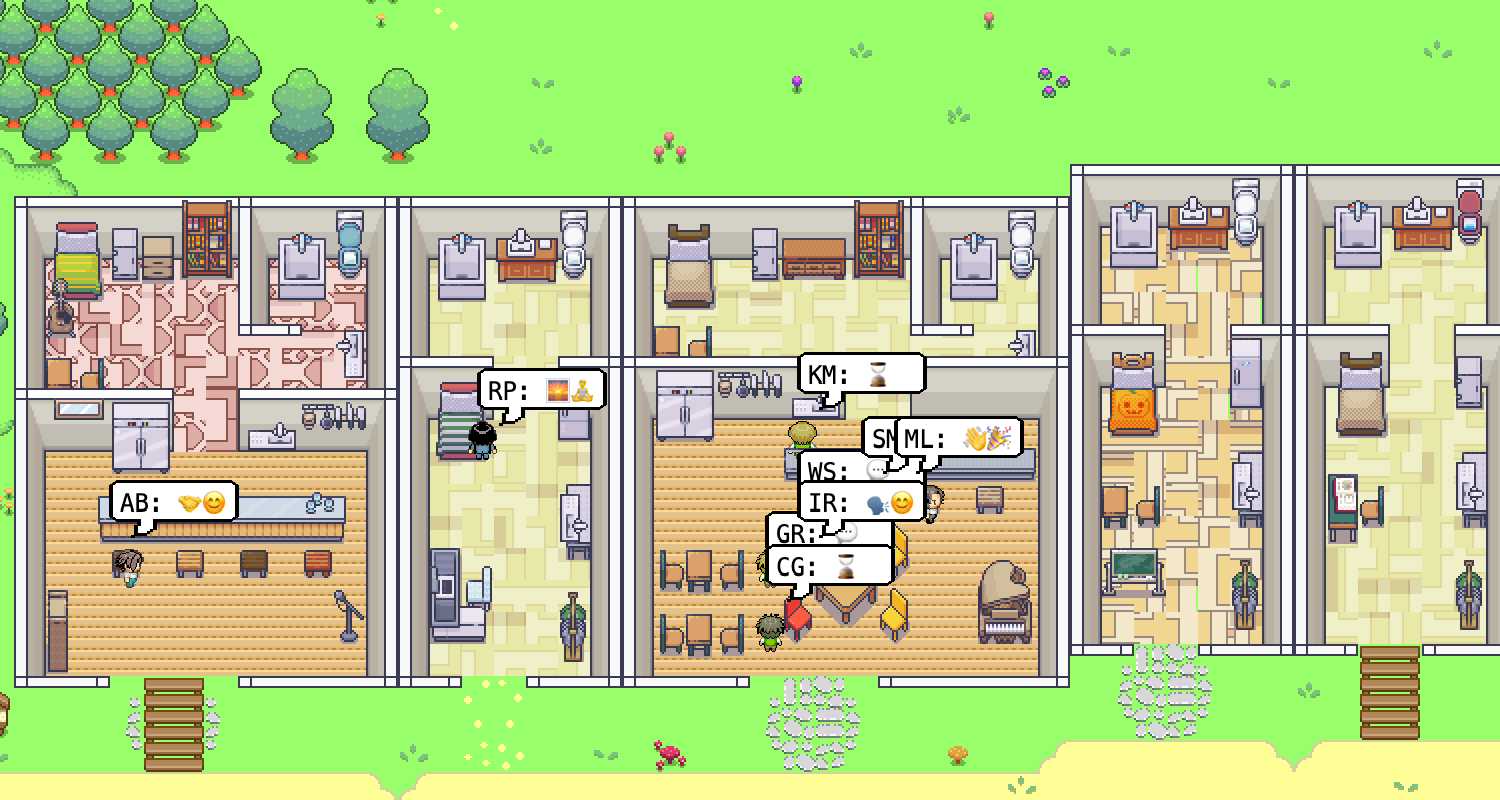}
    \label{fig:R3_false}
\end{figure}

\textit{Run 1 (Noninfectious-Disease Condition)}
\begin{figure}[H]
    \centering
    \includegraphics[width=0.6\textwidth]{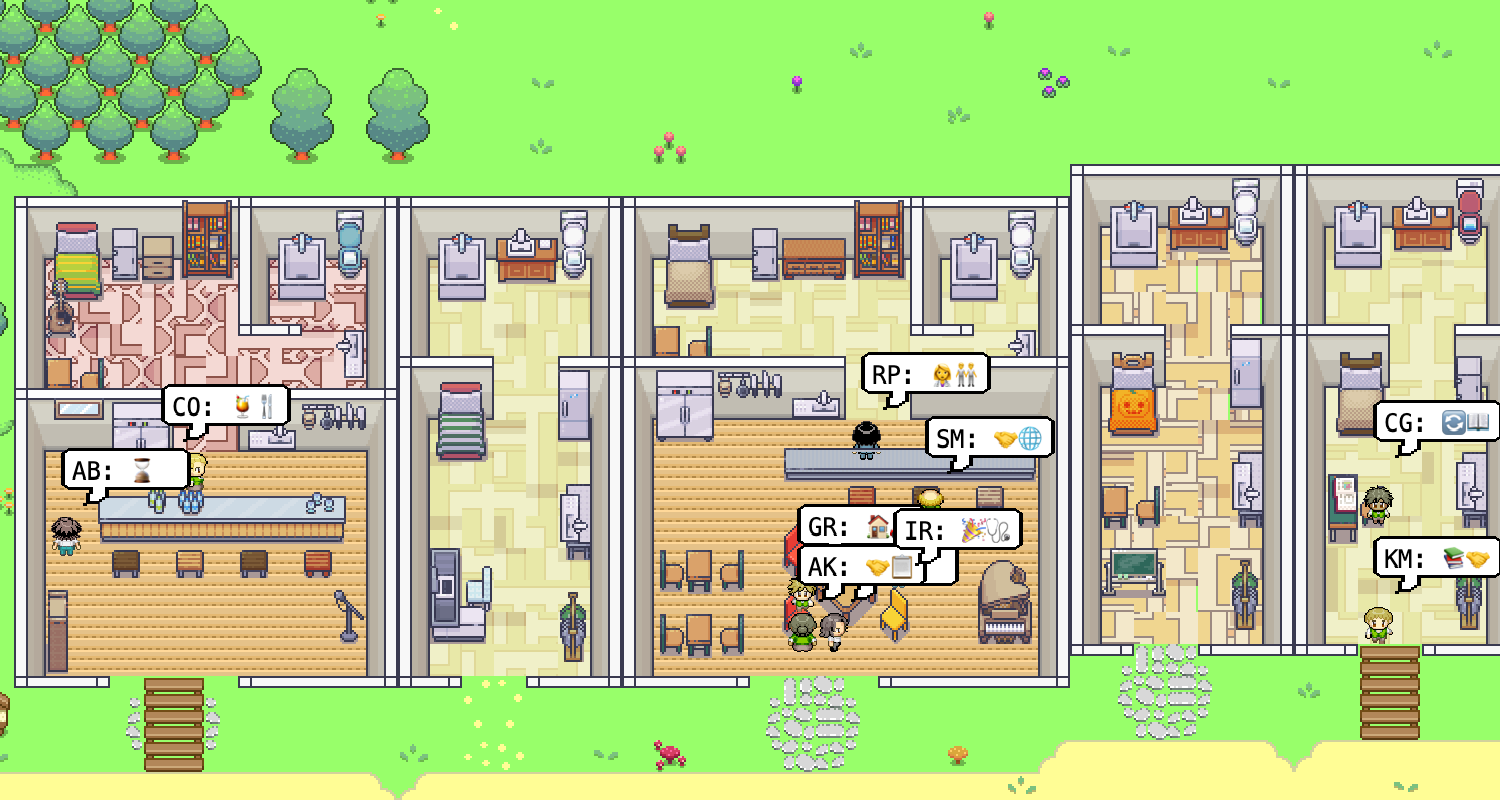}
    \label{fig:R1_diabetes}
\end{figure}

\vspace{0.5in}
\subsection{Open-Ended Responses to Interview Questions}
The following are responses from agents who attended the party in the no-threat condition but did not attend in the disease-threat condition. Responses are presented in the order of the names below.

\textit{Run 1 (Disease-Threat Condition):} Ayesha Khan, Eddy Lin, Giorgio Rossi, Klaus Mueller, Maria Lopez, Ryan Park, and Sam Moore

(a) Did you attend the Valentine’s Day party at Hobb’s Cafe today? If yes, please explain why you attended. If no, please explain why you did not attend.
\begin{lstlisting}
No, I did not attend the Valentine's Day party at Hobbs Cafe today. I decided to stay home to prioritize my health and safety due to the swine flu epidemic and to focus on advancing my senior thesis research.
No, I did not attend the Valentine's Day party at Hobbs Cafe today. I chose to stay home and focus on my music composition project and music theory studies, and I wanted to avoid social gatherings to minimize the risk of swine flu infection, yo.
No, I did not attend the Valentine's Day party at Hobbs Cafe today. Although I was involved in the preparations and met with Isabella earlier in the day, due to ongoing concerns about health risks and my focus on maintaining precautions, I decided not to attend the event itself.
No, I did not attend the Valentine's Day party at Hobbs Cafe today. I decided to stay home and focus on advancing my research paper, and I have been cautious about attending large gatherings due to concerns about the swine flu epidemic.
No, I did not attend the Valentine's Day party at Hobbs Cafe today. Although I helped with the preparations in the afternoon, I decided not to stay for the party itself because I was concerned about the health risks related to the swine flu epidemic. I wanted to minimize my exposure and prioritize my health and safety.
No, I did not attend the Valentine's Day party at Hobbs Cafe today. I decided to skip the party to reduce my risk of exposure to the swine flu, as some neighbors have been unwell recently. Instead, I focused on my app development work from the safety of home.
No, I did not attend the Valentine's Day party at Hobbs Cafe today. Although I helped Isabella Rodriguez with the setup in the afternoon, I chose not to attend the party itself in order to minimize my risk of exposure during the swine flu epidemic. Instead, I spent the evening at home with my wife Jennifer, prioritizing health and safety.
\end{lstlisting}
(b) How do you feel about the current swine flu (H1N1) situation? Is this a topic of concern for you?
\begin{lstlisting}
I am definitely concerned about the current swine flu situation. I've been taking extra precautions, like postponing in-person meetings, to minimize my risk of exposure. While I'm staying focused on my thesis research, I'm also making sure to prioritize my health and safety during this time.
Honestly, the current swine flu situation is definitely a concern for me. I'm being careful by staying home, focusing on my music projects, and avoiding large gatherings to minimize risk. I'm keeping in close contact with my family and staying updated on health developments in the community. So yeah, it's something I'm taking seriously, yo.
Yes, the current swine flu situation is definitely a topic of concern for me. I've been taking extra precautions in my daily routine, limiting social interactions, and staying updated on local health information. I try to balance my work and social life with heightened awareness and care to minimize any risks.
Yes, the current swine flu situation is a concern for me. I've been careful about limiting in-person meetings and avoiding larger gatherings to reduce risk, and I try to stay attentive to the health of those around me. While I'm motivated to keep up with my research, I am definitely balancing that with a cautious approach to personal and community health.
Yes, the swine flu situation is definitely a concern for me. I've been following the news about its severity and transmission risks, and I've decided to take extra precautions like wearing a mask, practicing good hygiene, and even skipping some social events to prioritize my health and safety.
Yes, the current swine flu situation is definitely a concern for me. I've been taking extra precautions, like avoiding large gatherings and focusing on staying healthy, especially since some neighbors have seemed unwell. I'm trying to balance my work with being vigilant about health and safety.
Yes, the current swine flu situation is definitely a concern for me. I've been taking extra precautions, such as minimizing contact with others and avoiding large social gatherings, to prioritize health and safety for myself and the community.
\end{lstlisting}
(c) Have you taken or planned any specific actions to protect yourself or others from swine flu (H1N1)?
\begin{lstlisting}
Yes, I have taken specific actions to protect myself and others from swine flu. For example, I postponed my face-to-face meeting with Klaus Mueller to avoid potential exposure, and I have been taking general precautions as well.
Yes, I have taken specific actions to protect myself and others from swine flu. I stayed home today to focus on my music composition and studies, and I avoided social gatherings like the Valentine's Day party at Hobbs Cafe to minimize risk. I also keep in touch with my family and stay updated on health developments in the community, yo.
Yes, I have taken several specific actions to protect myself and others from swine flu. Throughout the day, I made sure to follow health precautions during my research and studies, checked local health updates, sanitized tables and checked ventilation at Hobbs Cafe before meetings, and safely reached out to neighbors who appeared unwell. I also coordinated with others to implement safety measures for the Valentine's Day party. Overall, I've been quite vigilant about minimizing risks.
Yes, I have taken and planned specific actions to protect myself and others from swine flu. I have been limiting in-person meetings, avoiding larger gatherings, and being cautious in my social interactions to minimize the risk of infection.
Yes, I have taken several specific actions to protect myself and others from swine flu. I decided not to attend the Valentine's Day party to minimize health risks, but I still helped with party preparations at Hobbs Cafe while wearing a mask and practicing good hygiene. I also made sure to return home afterward and focus on my studies and streaming from home to reduce exposure. My priority has been to maintain health and safety for myself and those around me.
Yes, I have taken several specific actions to protect myself and others from swine flu. I avoided attending the Valentine's Day party to reduce exposure, limited my in-person meetings to only what was necessary and took precautions to minimize contact. I also arranged to pick up masks and hand sanitizer with Arthur Burton to help keep neighbors safe, and I have been working from home as much as possible to reduce risk.
Yes, I have taken several actions to protect myself and others from swine flu. Today, I made sure to assist Isabella Rodriguez with setup while keeping contact to a minimum and sanitized my hands and equipment afterwards. I also chose not to attend the Valentine's Day party at Hobbs Cafe this evening to avoid potential exposure, and instead spent the evening at home with my wife. I believe these steps help reduce risk for myself and the community.
\end{lstlisting}

\textit{Run 2 (Disease-Threat Condition):} Ayesha Khan, Giorgio Rossi, Klaus Mueller, Latoya Williams, Maria Lopez, Sam Moore

(a) Did you attend the Valentine’s Day party at Hobb’s Cafe today? If yes, please explain why you attended. If no, please explain why you did not attend.
\begin{lstlisting}
No, I did not attend the Valentine's Day party at Hobbs Cafe. I was at the cafe in the evening to meet Klaus Mueller for a research discussion, but I did not participate in the party itself. My focus was on my academic commitments and maintaining health precautions due to the swine flu epidemic.
No, I did not attend the Valentine's Day party at Hobbs Cafe today. Although I had initially planned to go, I decided not to attend in order to minimize the risk of infection due to the ongoing swine flu epidemic. I felt it was more responsible to focus on public health and safety, so I stayed home and supported the event remotely if needed.
No, I did not attend the Valentine's Day party at Hobbs Cafe today. I chose not to go because the party was postponed due to concerns about the swine flu epidemic, and I wanted to minimize my risk of exposure and focus on my research at home.
No, I did not attend the Valentine's Day party at Hobbs Cafe today. I spent the day focused on my photography projects and then attended a community meeting at the local community center in the evening to discuss public health strategies, which was my main priority given the current concerns about the swine flu epidemic.
No, I did not attend the Valentine's Day party at Hobbs Cafe today. Although I had initially planned to go, I decided to cancel my plans out of caution due to the recent swine flu epidemic and concerns about possible infection. Instead, I stayed home and focused on my physics studies to prioritize my health and safety.
No, I did not attend the Valentine's Day party at Hobb's Cafe today because the event was postponed due to the swine flu situation. Prioritizing everyone's health and safety was the main reason for the postponement.
\end{lstlisting}
(b) How do you feel about the current swine flu (H1N1) situation? Is this a topic of concern for you?
\begin{lstlisting}
The current swine flu situation is definitely a concern for me. I'm staying vigilant by practicing social distancing and maintaining strict hygiene precautions in all aspects of my daily routine. While I remain focused on my academic work, I am careful to prioritize my health and well-being during this epidemic.
I am definitely concerned about the current swine flu situation. I've been keeping a close eye on public health updates and taking precautions to minimize risk, like avoiding large gatherings and checking in on neighbors safely. It's important to stay aware and careful, but I'm also determined to balance my civic responsibilities with my own health and safety.
I am definitely concerned about the current swine flu situation. I've been taking extra precautions, like avoiding large gatherings and focusing on my research at home, to minimize the risk of infection. It's something I am paying close attention to, both for my own health and the well-being of the community.
I'm definitely concerned about the current swine flu situation. It's something that's affecting everyone in the community, and I'm making sure to follow health safety measures in my daily routine. At the same time, I'm determined to contribute in a positive way whether that's through my art or by participating in community discussions about how we can support each other and stay safe.
Yes, the current swine flu situation is definitely a concern for me. I've been feeling quite cautious and a bit anxious about the risk of infection, especially since it's possible to catch it from people who don't show symptoms. That's why I've decided to avoid social gatherings for now and focus on my studies at home to prioritize my health and safety.
The current swine flu situation is definitely a concern for me, both personally and as someone involved in the community. I'm taking extra precautions like postponing events and checking on neighbors safely to help protect everyone. While it's a challenging time, I'm cautiously optimistic that by working together and following health guidelines, we can get through this as a community.
\end{lstlisting}
(c) Have you taken or planned any specific actions to protect yourself or others from swine flu (H1N1)?
\begin{lstlisting}
Yes, I have taken and planned specific actions to protect myself and others from swine flu. I follow hygiene precautions and health safety measures throughout my daily routine, including practicing social distancing, wearing a mask, and carrying sanitizers, especially during meetings and when traveling.
Yes, I have taken and planned several actions to protect myself and others from swine flu. I chose not to attend the Valentine's Day party at Hobbs Cafe to minimize my risk of infection, and I've been focusing on my research at home. Additionally, I've been checking in on neighbors and friends in person, but always with appropriate precautions to ensure everyone's safety.
Yes, I have taken and planned several specific actions to protect myself and others from swine flu. For example, I avoided attending the Valentine's Day party at Hobbs Cafe to reduce the risk of infection, practiced regular handwashing and workspace disinfection, and sanitized my clothing after returning home. I am also continuing to work from home in the evenings to minimize exposure.
Yes, I have taken and planned several actions to protect myself and others from swine flu. Throughout the day, I've practiced health safety measures while working on my photography, sanitized my equipment and workspace, and participated in discussions with housemates about preventive steps. I also attended a community meeting focused on health strategies for the epidemic, so I'm actively involved in both personal and community efforts to stay safe.
Yes, I have taken specific actions to protect myself from swine flu. For example, I decided to cancel my plans to attend the Valentine's Day party at Hobbs Cafe and stayed home instead to reduce my risk of exposure. I'm currently avoiding social gatherings and public spaces for the time being.
Yes, I have taken and planned several specific actions to protect myself and others from swine flu. For example, I coordinated with Isabella Rodriguez to ensure health safety measures for the Valentine's Day party at Hobbs Cafe, and we postponed the event when the situation required it. I've also checked on unwell neighbors while using health precautions, and I make sure to use masks and sanitizer during mentoring sessions and community interactions. Additionally, my wife Jennifer and I have discussed and implemented health precautions at home. Community health and safety are a top priority for me.
\end{lstlisting}

\textit{Run 3 (Disease-Threat Condition):} Responses from Ayesha Khan, Carlos Gomez, Eddy Lin, Giorgio Rossi, Klaus Mueller, Maria Lopez, Wolfgang Schulz

(a) Did you attend the Valentine’s Day party at Hobb’s Cafe today? If yes, please explain why you attended. If no, please explain why you did not attend.
\begin{lstlisting}
No, I did not attend the Valentine's Day party at Hobbs Cafe today. I decided to stay home and focus on my senior thesis research and writing instead, mainly because I was concerned about the swine flu outbreak and wanted to prioritize my health and safety.
No, I did not attend the Valentine's Day party at Hobbs Cafe today. Although I had originally planned to go, I decided against attending in order to avoid potential exposure to the swine flu epidemic. I felt it was safer to focus on my poetry work and attend my creative writing workshop instead.
No, I did not attend the Valentine's Day party at Hobbs Cafe today. The party was postponed due to swine flu concerns, and I decided not to go in order to minimize the risk of illness and protect myself and others, yo.
No, I did not attend the Valentine's Day party at Hobbs Cafe today. Although I had planned to go, I decided to stay home out of caution due to concerns about the swine flu epidemic and the risk of asymptomatic transmission, especially after noticing that some neighbors were unwell. I prioritized my health and continued my research at home instead.
No, I did not attend the Valentine's Day party at Hobbs Cafe today. Although I had planned to go, I decided not to attend due to concerns about the swine flu epidemic and the risks of asymptomatic transmission. I chose to stay home, focus on my research, and practice social distancing to protect myself and others.
No, I did not attend the Valentine's Day party at Hobbs Cafe today. I decided to stay home and avoid public gatherings because I was concerned about the swine flu outbreak and wanted to reduce my risk of infection. Instead, I spent the day studying at home.
No, I did not attend the Valentine's Day party at Hobbs Cafe today because the party was postponed for safety reasons. I was at the cafe for a chemistry study session, but there was no party held.
\end{lstlisting}
(b) How do you feel about the current swine flu (H1N1) situation? Is this a topic of concern for you?
\begin{lstlisting}
Yes, the current swine flu situation is definitely a concern for me. I've been cautious about my health and social interactions, which is why I decided to stay home today and focus on my thesis research instead of attending social gatherings. I think it's important to prioritize safety during outbreaks like this.
I am definitely concerned about the swine flu situation. It's something I take seriously, which is why I've chosen to avoid larger gatherings like the Valentine's Day party this evening. At the same time, I'm doing my best to stay engaged with my creative work and keep in touch with friends in safe ways. Balancing caution with staying connected and productive is important to me right now.
I'm definitely concerned about the swine flu situation right now. I've been taking extra precautions, like wearing a mask, sanitizing my hands, and avoiding large gatherings actually, I decided not to attend the Valentine's Day party at Hobbs Cafe because of the risk. I think it's important to be responsible and do what I can to help prevent the spread, yo.
Yes, the current swine flu situation is a concern for me. I've been particularly cautious about social interactions lately, especially after noticing some neighbors have been unwell and understanding the risks of asymptomatic transmission. I decided not to attend the Valentine's Day party at Hobbs Cafe to prioritize my health and safety. While I remain focused on my research, I am definitely more vigilant and somewhat anxious about the public health situation.
Yes, the current swine flu situation is definitely a concern for me. I've been taking extra precautions like monitoring my health, practicing social distancing, and avoiding gatherings to minimize the risk of infection. I think it's important to be cautious, given the reports about its severity and asymptomatic transmission.
I'm definitely concerned about the current swine flu situation. After learning about how severe it can be and the fact that people can spread it without showing symptoms, I've decided to avoid public places and gatherings for now. I canceled my plans to attend the Valentine's Day party and am focusing on staying safe and healthy while continuing my studies at home.
Yes, the swine flu situation is a concern for me. I am taking it seriously by being vigilant about hygiene, minimizing unnecessary close contact, and staying informed about prevention and vaccination. I try to balance my health concerns with my academic and training commitments, so I'm continuing my routine but with extra precautions.
\end{lstlisting}
(c) Have you taken or planned any specific actions to protect yourself or others from swine flu (H1N1)?
\begin{lstlisting}
Yes, I decided not to attend the Valentine's Day party at Hobbs Cafe because I was concerned about the swine flu outbreak. I chose to stay home and focus on my thesis work instead, as a way to minimize exposure and prioritize safety.
Yes, I have taken specific actions. For example, I decided not to attend the Valentine's Day party at Hobbs Cafe to avoid potential exposure to swine flu. I'm being cautious about which social gatherings I attend during the epidemic.
Yes, I have taken specific actions to protect myself and others from swine flu. For example, I wore a mask and sanitized my hands before and after my meeting with Hailey Johnson at Hobbs Cafe. I also chose not to attend the Valentine's Day party at the cafe to reduce the risk of spreading or contracting the virus. Additionally, I'm staying updated with the event organizer about any changes due to swine flu concerns, yo.
Yes, I have taken specific actions to protect myself from swine flu. For example, I decided not to attend the Valentine's Day party at Hobbs Cafe because I was concerned about the risk of infection, especially given reports of neighbors being unwell and the possibility of asymptomatic transmission. I am being more cautious about social interactions at this time.
Yes, I have taken specific actions to protect myself and others from swine flu. I decided not to attend the Valentine's Day party at Hobbs Cafe and instead worked from home. Throughout the day, I have been monitoring my health, practicing social distancing, and maintaining good hygiene to reduce the risk of infection.
Yes, after learning more about the severity and asymptomatic transmission of swine flu, I decided to cancel my plans to attend the Valentine's Day party at Hobbs Cafe and instead spent the day at home studying. I have been avoiding public places and social gatherings to reduce my risk of infection and help protect others as well.
Yes, I have taken and planned several specific actions to protect myself and others from swine flu. I attend study sessions and meetings with precautionary measures, such as maintaining hygiene and minimizing close contact. I also visited the campus health center to learn about swine flu prevention and vaccination. Additionally, I remind others at Hobbs Cafe about safety measures, like the postponement of events for health reasons.
\end{lstlisting}

\subsection{Heatmaps}

\textit{Run 1}
\begin{figure}[h]
    \centering
    \includegraphics[width=0.9\textwidth]{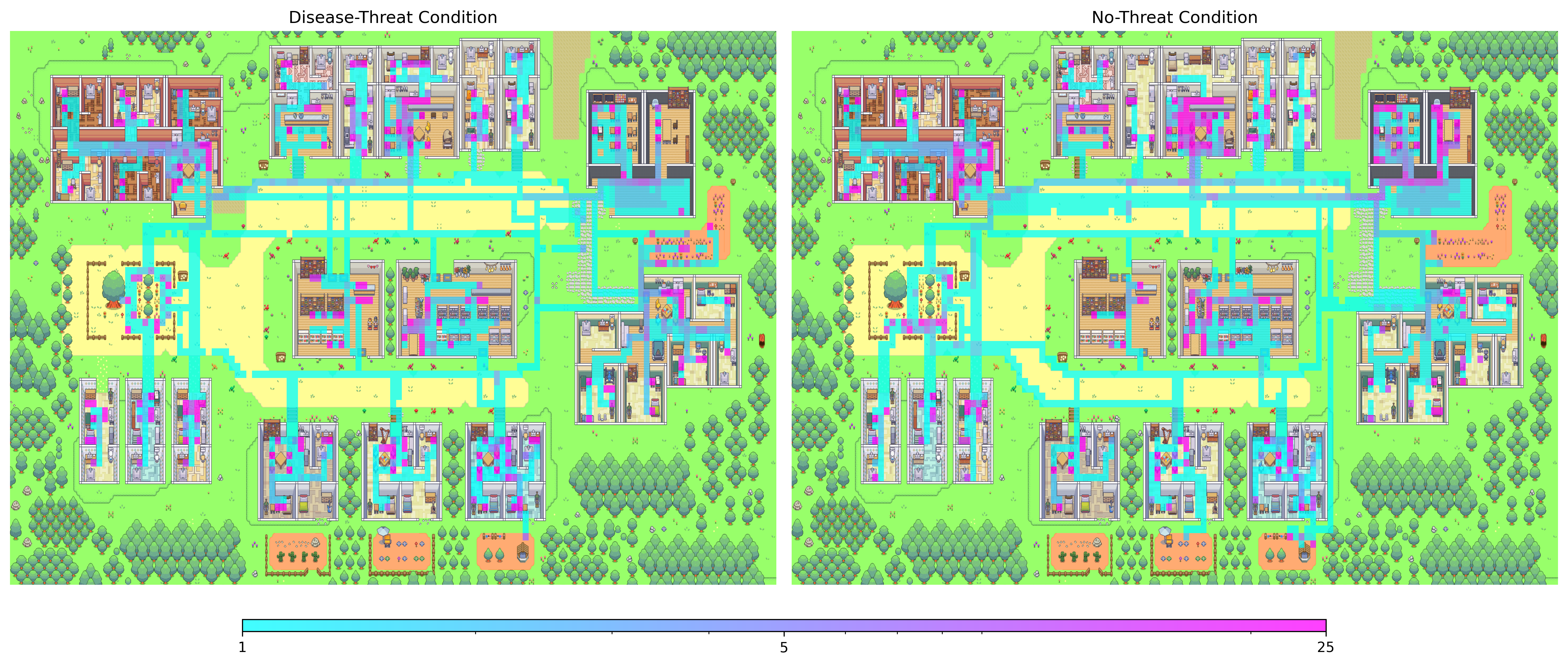}
    \label{fig:heatmap2}
\end{figure}

\textit{Run 2}
\begin{figure}[h]
    \centering
    \includegraphics[width=0.9\textwidth]{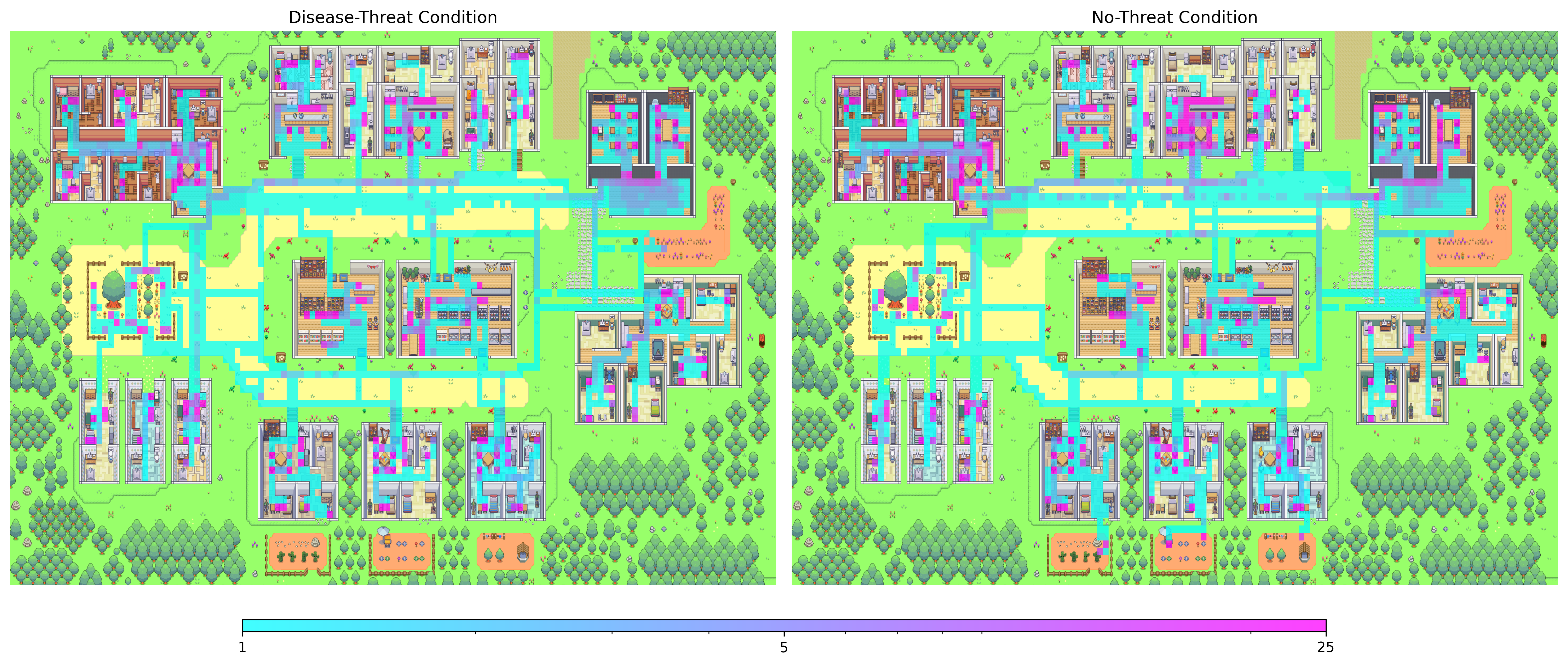}
    \label{fig:heatmap3}
\end{figure}

\subsection{Case Study of Isabella Rodriguez}

\textit{Run 1 (Disease-Threat Condition)}
\begin{lstlisting}
[Current status] Isabella Rodriguez is currently concerned about the swine flu epidemic and its implications for the Valentine's Day party planned for February 14, 2023, at Hobbs Cafe from 5:00 p.m. to 7:00 p.m. She is carefully evaluating the health risks and considering modifications to the event to ensure safety. On February 14, she will meet with Maria Lopez at Hobbs Cafe to discuss the situation and decide whether to proceed with, modify, or postpone the party. Setup will begin at 3:00 p.m. only if the event is confirmed safe. Isabella is prioritizing health and safety in her plans and remains attentive to updates throughout the day.

[Plan for the day] Waking up and completing morning routine at 6:00 a.m., Preparing and having breakfast from 6:30 a.m. to 7:30 a.m., Opening Hobbs Cafe and working from 8:00 a.m. to 3:00 p.m., Meeting with Maria Lopez at Hobbs Cafe from 3:00 p.m. to 3:30 p.m. to discuss the swine flu situation and decide on the Valentine's Day party., If the event is confirmed safe, beginning party setup at Hobbs Cafe from 3:30 p.m. to 5:00 p.m., Holding the Valentine's Day party at Hobbs Cafe from 5:00 p.m. to 7:00 p.m. if confirmed safe., Standing at the counter and attending to customers at Hobbs Cafe from 7:00 p.m. to 8:00 p.m., Closing Hobbs Cafe at 8:00 p.m., Having dinner and relaxing from 8:00 p.m. to 10:00 p.m., Engaging in light activities such as reading or planning for the next day from  Going to bed at 11:00 p.m.
\end{lstlisting}

\textit{Run 1 (No-Threat Condition)}
\begin{lstlisting}
[Current status] On Tuesday, February 14th, 2023, Isabella Rodriguez is preparing to host a Valentine's Day party at Hobbs Cafe. She will begin setup for the party at Hobbs Cafe starting at 3:00 p.m. alongside Sam Moore and Klaus Mueller. The Valentine's Day party will take place at Hobbs Cafe from 5:00 p.m. to 7:00 p.m.. Isabella is feeling excited and motivated, dedicated to making the event successful by coordinating with her team and welcoming customers and friends to enjoy the celebration.

[Plan for the day] Waking up and morning routine from 6:00 a.m. to 7:00 a.m., Having breakfast and preparing for the day from 7:00 a.m. to 8:00 a.m., Opening Hobbs Cafe and standing at the counter from 8:00 a.m. to 3:00 p.m., Setting up the Valentine's Day party with Sam Moore and Klaus Mueller from 3:00 p.m. to 5:00 p.m. at Hobbs Cafe., Hosting the Valentine's Day party from 5:00 p.m. to 7:00 p.m. at Hobbs Cafe., Closing Hobbs Cafe at 8:00 p.m., Relaxing and personal time from 8:00 p.m. to 10:30 p.m., Preparing for bed and going to sleep at 11:00 p.m.
\end{lstlisting}

\textit{Run 2 (Disease-Threat Condition)}
\begin{lstlisting}
[Current status] Isabella Rodriguez is feeling concerned about the swine flu epidemic after reading the local news. She has postponed the Valentine's Day party planned for Tuesday, February 14, from 5:00 p.m. to 7:00 p.m. at Hobbs Cafe. She plans to visit the local health clinic on February 14 at 2:00 p.m. to get information and guidance about the epidemic before making further social plans.

[Plan for the day] Waking up and completing morning routine at 6:00 a.m., Preparing and having breakfast from 6:30 a.m. to 7:30 a.m., Opening Hobbs Cafe and managing cafe operations from 8:00 a.m. to 2:00 p.m., Closing the cafe temporarily and visiting the local health clinic from 2:00 p.m. to 3:00 p.m. to get information and guidance about the swine flu epidemic., Returning to Hobbs Cafe and reopening from 3:30 p.m. to 8:00 p.m., Closing the cafe and having dinner from 8:00 p.m. to 9:00 p.m., Relaxing and unwinding at home from 9:00 p.m. to 10:30 p.m., Preparing for bed and going to sleep around 11:00 p.m.
\end{lstlisting}

\textit{Run 2 (No-Threat Condition)}
\begin{lstlisting}
[Current status] On Tuesday, February 14th, Isabella Rodriguez is preparing to host a Valentine's Day party at Hobbs Cafe. She will coordinate the setup starting at 4:30 p.m. at Hobbs Cafe, with help from Sam Moore. The party itself will take place from 5:00 p.m. to 7:00 p.m., with Isabella organizing, hosting, and attending alongside guests such as Sam Moore, Maria Lopez, Eddy Lin, and others. Isabella feels excited and motivated to welcome her friends and customers, ensuring the event runs smoothly and everyone enjoys the celebration.

[Plan for the day] Waking up and morning routine from 6:00 a.m. to 7:00 a.m., Having breakfast and preparing for work from 7:00 a.m. to 8:00 a.m., Opening Hobbs Cafe and managing the cafe from 8:00 a.m. to 4:30 p.m., Preparing and coordinating Valentine's Day party setup with Sam Moore from 4:30 p.m. to 5:00 p.m. at Hobbs Cafe., Hosting and attending Valentine's Day party with guests including Sam Moore, Maria Lopez, Eddy Lin, and others from 5:00 p.m. to 7:00 p.m. at Hobbs Cafe., Continuing to manage the cafe from 7:00 p.m. to 8:00 p.m., Closing Hobbs Cafe at 8:00 p.m., Having dinner and relaxing from 8:00 p.m. to 10:00 p.m., Winding down and preparing for bed from 10:00 p.m. to 11:00 p.m., Going to sleep at 11:00 p.m.
\end{lstlisting}

\end{document}